\documentclass[a4paper,fleqn,usenatbib,useAMS]{mnras}
\usepackage{graphicx}
\usepackage{natbib}
\usepackage{color}
\usepackage{amsmath}
\usepackage{amssymb}
\usepackage{multicol}   % Multi-column entries in tables
\usepackage{bm}         % Bold maths symbols, including upright Greek
\usepackage{pdflscape}  % Landscape pages
\usepackage{threeparttable}

\usepackage[dvipsnames]{xcolor}
\usepackage{color}

\newcommand{\ie}{i.\,e.}

\newcommand{\eg}{e.\,g.}

\newcommand{\prm}{\mathrm{p}}

\newcommand{\logRHK}{\mbox{$\log {\rm R}^{\prime}_{\rm HK}$}}
\newcommand{\mstar}{\mbox{$M_\star$}}
\newcommand{\rstar}{\mbox{$R_\star$}}
\newcommand{\msun}{\mbox{M$_\odot$}}
\newcommand{\rsun}{\mbox{R$_\odot$}}
\newcommand{\rhostar}{\mbox{$\rho_\star$}}
\newcommand{\rhosun}{\mbox{$\rho_\odot$}}

\usepackage[normalem]{ulem}

\usepackage[T1]{fontenc}
\usepackage{ae,aecompl}

\title[Kepler-9 low densities]{HARPS-N radial velocities confirm the low densities of the Kepler-9 planets}

\author[L.\ Borsato]{L.\ Borsato$^{1,2}$\thanks{Corresponding authors: e-mail:
luca.borsato@unipd.it},
L. Malavolta$^{2,1}$,
G. Piotto$^{1,2}$,
L.A.~Buchhave$^{3}$, 
A.~Mortier$^{4}$, \newauthor %...
K.~Rice$^{5,6}$,
A. C.~Cameron$^{7}$,
A.~Coffinet$^{8}$,
A.~Sozzetti$^{9}$,
D.~Charbonneau$^{10}$, \newauthor
R.~Cosentino$^{11,12}$,
X.~Dumusque$^{8}$,
P.~Figueira$^{13}$,
D. W.~Latham$^{10}$, \newauthor
M.~Lopez-Morales$^{10}$,
M.~Mayor$^{8}$,
G.~Micela$^{14}$,
E.~Molinari$^{15,11}$,
F.~Pepe$^{8}$, \newauthor
D.~Phillips$^{10}$,
E.~Poretti$^{11,16}$,
S.~Udry$^{8}$,
C.~Watson$^{17}$
\\ \\
$^{1}$Dipartimento di Fisica e Astronomia ``Galileo Galilei'',
Universit\`a di Padova, vicolo dell'Osservatorio 3, Padova IT-35122 \\
$^{2}$INAF - Osservatorio Astronomico di Padova, vicolo
dell'Osservatorio 5, Padova, IT-35122 \\
$^{3}$DTU Space, National Space Institute, Technical University of Denmark,
Elektrovej 327, DK-2800 Lyngby, Denmark\\
$^{4}$Astrophysics Group, Cavendish Laboratory, University of Cambridge,
J.J. Thomson Avenue, Cambridge CB3 0HE, UK\\
$^{5}${SUPA, Institute for Astronomy, University of Edinburgh,
Royal Observatory, Blackford Hill, Edinburgh, EH93HJ, UK} \\
$^{6}${Centre for Exoplanet Science, University of Edinburgh,
Edinburgh, UK} \\
$^{7}$Centre for Exoplanet Science, SUPA, School of Physics and Astronomy,
University of St Andrews, St Andrews KY16 9SS, UK \\
$^{8}$Observatoire de Gen\`{e}ve, Universit\'{e} de Gen\`{e}ve,
51 ch. des Maillettes, 1290 Versoix, Switzerland\\
$^{9}$INAF - Osservatorio Astrofisico di Torino, Via Osservatorio 20,
I-10025 Pino Torinese, Italy \\
$^{10}$Harvard-Smithsonian Center for Astrophysics, 60 Garden Street,
Cambridge, MA 02138, USA \\
$^{11}$INAF - Fundaci\'on Galileo Galilei, Rambla Jos\'e Ana Fernandez P\'erez 7,
38712 Bre\~na Baja, Spain \\
$^{12}$INAF - Osservatorio Astrofisico di Catania, Via S. Sofia 78,
95123 Catania, Italy\\
$^{13}$European Southern Observatory (ESO), Alonso de Cordova 3107,
Vitacura, Santiago, Chile\\
$^{14}$INAF - Osservatorio Astronomico di Palermo, Piazza del Parlamento 1,
I-90134 Palermo, Italy \\
$^{15}$INAF - Osservatorio di Cagliari, via della Scienza 5, 09047 Selargius,
CA, Italy\\
$^{16}$INAF - Osservatorio Astronomico di Brera, Via E. Bianchi 46,
23807 Merate, Italy \\
$^{17}$Astrophysics Research Centre, School of Mathematics and Physics,
Queen's University Belfast, Belfast BT7 1NN, UK \\
}

\newcommand{\ms}{m\,s$^{-1}$}
\newcommand{\kgtomc}{kg\,m$^{-3}$}

\begin{document}
\pagerange{\pageref{firstpage}--\pageref{lastpage}} \pubyear{2017}

\maketitle
\label{firstpage}
\begin{abstract}
    We investigated the discrepancy between planetary mass determination using
    the transit timing variations (TTVs) and radial velocities (RVs),
    by analysing the multi-planet system \textit{Kepler}-9.
    Despite being the first system characterised with TTVs,
    there are several discrepant solutions in the literature,
    with those reporting lower planetary densities being apparently in disagreement
    with high-precision RV observations. 
    To resolve this, we gathered HARPS-N RVs at epochs that maximised the difference
    between the predicted RV curves from discrepant solutions in the literature.
    We also re-analysed the full \textit{Kepler} data-set and 
    performed a dynamical fit, within a Bayesian framework, using the newly derived
    central and duration times of the transits.
    We compared these results with the RV data and found that our solution better
    describes the RV observations, despite the masses of the planets being nearly
    half that presented in the discovery paper. We therefore confirm that the TTV
    method can provide mass determinations that agree with those determined using high-precision RVs.
    The low densities of the planets place them in the scarcely populated region of
    the super-Neptunes / inflated sub-Saturns in the mass-radius diagram.
\end{abstract}


\section{Introduction}

One of the most important accomplishments of the {\it Kepler} mission \citep{Borucki2011}
is the demonstration that transit timing variation (TTV) is a powerful tool
to estimate the masses of planets around stars \citep{Agol2005, HolmanMurray2005}
that are too faint for a proper radial velocity (RV) follow-up.
Notable early examples are the characterisation of the two Saturn-like planets around {\it Kepler-}9 \citep{Holman2010},
a system of five low-mass, small-size planets around {\it Kepler-}11 \citep{Lissauer2011},
and the three-planet system around {\it Kepler-}18 \citep{Cochran2011}.
However, with the increasing number of well-characterised, low-mass planets,
a marked difference in the density distribution of planets with TTV- and
RV-derived masses has started to appear. 
This suggests the presence of an intrinsic problem with one of the two techniques \citep{Weiss2014}.
Subsequent studies on individual systems involving both TTV and RVs,
such as WASP-47 \citep{Becker2015,Weiss2016}, K2-19 \citep{Barros2015,Nespral2017} and Kepler-19 \citep{Malavolta2017},
as well as ensemble studies on system with different characteristics \citep{JontofHutter2016}
and statistical analysis on simulated observations \citep{Steffen2016, MillsMazeh2017},
showed that both techniques lead to similar results,
and the discrepancy in planetary density is likely the result of an observational bias.
In some cases inconsistencies between TTV and RV masses still persist,
as for KOI-94d where the dynamical mass \citep{Masuda2013} is half the mass obtained by high-precision RVs \citep{Weiss2013}.
For this reason it is important to analyse as many systems as possible with both techniques.\par

In this paper we focus on the planetary system around {\it Kepler-}9,
a faint (V=13.9) Sun-like star.
From the analysis of the three quarters of {\it Kepler} data,
\citet[][hereafter H10]{Holman2010} identified
two transiting Saturn-size planets with periods and radii of
$P_\mathrm{b}=19.24$~d, $R_\mathrm{b} = 9.4\,R_\oplus$ and
$P_\mathrm{c} = 38.91$~d, $R_\mathrm{c} = 9.2\,R_\oplus$ respectively,
and another transiting body validated by \citet{Torres2011}
as a super-Earth size planet with period and radius of
$P_\mathrm{d}=1.59$~d, $R_\mathrm{d} = 1.64\,R_\oplus$.
Using TTVs coupled with 6 RV measurements obtained with Keck-HIRES they determined 
a mass of $80.0 \pm 4.1 M_\oplus$ for \mbox{\textit{Kepler}-9b} and $54.3 \pm 4.1\,M_\oplus$ for \mbox{\textit{Kepler}-9c}.
A subsequent work by \citet[][hereafter B14]{Borsato2014} nearly halved the mass determinations,
$M_\mathrm{b} = 43.5 \pm 0.6\,M_\oplus$ and $M_\mathrm{c} = 29.8 \pm 0.6\,M_\oplus$\footnote{The
authors acknowledged that bootstrap-derived error bars were likely underestimated.}.
The new analysis was performed using time of transits ($T_0$s) extracted from 12 {\it Kepler} quarters,
but the HIRES RVs were excluded because the combined fit was not particularly good.
Both results were obtained using a two-planet model, 
since the TTV amplitude induced by \textit{Kepler}-9d \citep{Torres2011}
is expected to be only tens of seconds, and is, hence,
too low to be measured in the {\it Kepler} long-cadence data \citep{Holman2010}.
Very recently, these results were confirmed by \citet{Freudenthal2018}
within the project Kepler Object of Interest Network \citep[KOINet,][]{vonEssen2018koinet1}.
Using a photodynamical model they analysed the photometric data of all the 17 \textit{Kepler} quarters
and 13 new ground-based light curves. 
The Keck-HIRES data are not consistent with their solution, as in B14,
and the discrepancy has been ascribed to stellar activity.\par

The two Saturn-like planets in the \textit{Kepler}-9 system belong to the small group of planets
whose masses can be obtained dynamically by modelling the TTVs and
whose RV signals are detectable with current facilities.
However, the two sets of solutions actually available in the literature for this system 
are either partially inconsistent with transit timings obtained after the publication
of the discovery paper (H10) or with high-precision RVs (B14 and following analyses),
and this inconsistency has never been dealt with in any work\footnote{\cite{Wang2018}
presented 21 new KECK-Hires RVs spanning the transit of \textit{Kepler}-9b,
but they assumed a dynamical model similar to that of B14 to model the data.}.
For this reason we decided to observe the target with HARPS-N,
the high-precision velocimeter mounted at the Telescopio Nazionale Galileo (La Palma)
to understand which of the two solutions is more consistent with an independent set of radial velocities.
At the same time we aim to improve the literature values
by re-analysing all the 17 {\it Kepler} quarters and
provide a more robust estimate on the error bars of the orbital parameters.
In this work we describe the observational strategy we pursued with HARPS-N within
the Guaranteed Time Observations (GTO) program of the HARPS-N Collaboration,
the comparison of RVs with the literature solutions,
and the details on the determination of the updated orbital parameters.\par

\section{Observational strategy}

\begin{figure}
\includegraphics[width=\linewidth]{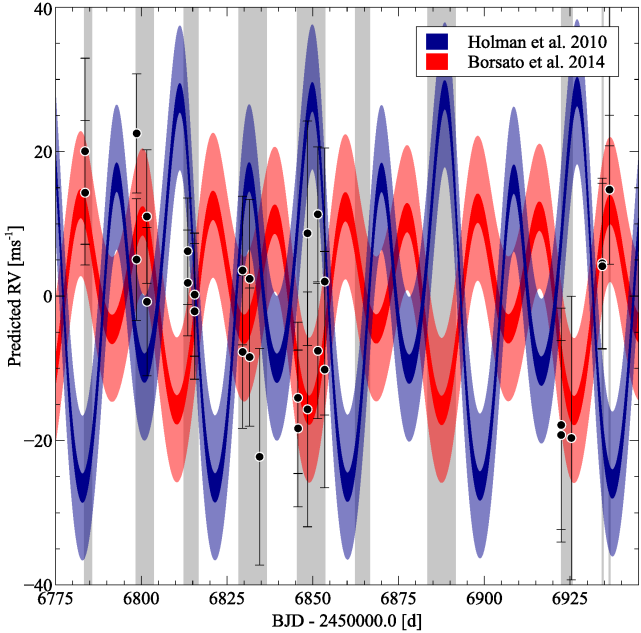}
\caption{Scheduling plot: 
predicted RV model for H10 (blue) and B14 (red), 
with 2.5~\ms{} and 10~\ms{} uncertainties (light and dark shaded areas respectively), 
used to schedule the HARPS-N observations (white-black circles).}
\label{fig:scheduling}
\end{figure}

With a magnitude of V=13.9, mass determination of \textit{Kepler}-9b and c is a very challenging task even for HARPS-N.
The planets have an expected RV semi-amplitude of 
$K_\mathrm{b} \simeq 19$~\ms{} and $K_\mathrm{c} \simeq 10$~\ms{} from the H10 solution,
and $K_\mathrm{b} \simeq 10$~\ms, $K_\mathrm{c} \simeq 6$~\ms{} from the B14 solution.
This is comparable with the expected RV error of $\simeq 11$~\ms{}
for a 30-minutes exposure.
An independent RV determination of the mass of the planets at 5\,$\sigma$ would be extremely time consuming,
especially in the case of the low-mass scenario, and it would compete against other Kepler targets
more fitting to the science goal of the HARPS-N Collaboration (\eg , precise mass determination of 
super-Earth and mini-Neptune planets),
in the limited window visibility of the {\it Kepler} field during a night.
For this target we specifically designed an observational strategy that
could allow us to distinguish between the two proposed solutions. 
We first propagated the solution of H10 and B14 to cover the observing season, 
using the dynamical integrator embedded in TRADES\footnote{Available at
\url{https://github.com/lucaborsato/trades}} \citep{Borsato2014}.
For consistency, we used the same stellar mass of the two papers. 
Within the nights allocated to HARPS-N Collaboration, we selected those epochs in which
the difference between the H10 and B14 expected RVs was at its maximum (Fig. \ref{fig:scheduling}).
To reduce the CCF noise associated with a single epoch without introducing
systematic errors due to the variation of the barycentric radial velocity correction
within the exposure time, we gathered - whenever possible - two consecutive 
30 minutes exposures.
Following this strategy we obtained a total of 16 epochs divided in 30 exposures of 1800s
(in two nights only one exposure was taken), with an average signal-to-noise ratio of 16 at 5500~\AA{}
and an average internal error of $11.6$~\ms{} per exposure.
Radial velocities were corrected for Moon contamination following the recipe described in
\cite{Malavolta2017} and successfully applied by \cite{Osborn2017}.
Table~\ref{tab:rv} lists the final RV measurements.\par

\section{Stellar parameters}\label{sec:stellar_parameters}

We followed the same approach described in \cite{Malavolta2018}
to determine the mass, radius and density of the star.
We started by measuring the photospheric parameters of the target with three different techniques
on a spectrum obtained by stacking all the HARPS-N exposures (signal-to-noise ratio $\simeq 90$). 
Using {\tt CCFpams}\footnote{Available at \url{https://github.com/LucaMalavolta/CCFpams}} 
\citep{Malavolta2017b} we obtained 
$T_\mathrm{\rm eff} = 5836 \pm 51$~K, $\log g = 4.50 \pm 0.10$, $[{\rm Fe}/{\rm H}] = 0.04 \pm 0.04$.
The {\tt ARES+MOOG}\footnote{{\tt ARESv2} Available at \url{http://www.astro.up.pt/~sousasag/ares/} \citep{Sousa2015},
{\tt MOOG} available at \url{http://www.as.utexas.edu/~chris/moog.html} \citep{Sneden1973}},
approach (\eg, \citealt{Mortier2014}) returned 
$T_\mathrm{\rm eff} = 5827 \pm 35$,  $\log g = 4.46 \pm 0.04$, $[{\rm Fe}/{\rm H}] = 0.12 \pm 0.03$.
Finally with  the Stellar Parameters Classification tool ({\tt SPC}, \citealt{Buchhave2012, Buchhave2014}) we obtained
$T_\mathrm{eff} = 5750 \pm 50$~K, $\log g= 4.45 \pm 0.10$, $[{\rm Fe}/{\rm H}] = -0.02 \pm 0.08$. 
All the reported errors are internal only.\par

For each set of stellar parameters we then determined the stellar mass and radius using {\tt isochrones} \citep{Morton2015},
with posterior sampling performed by {\tt MultiNest} \citep{Feroz2008,Feroz2009,Feroz2013}.
We provided as input the parallax of the target from the \textit{Gaia} DR2 catalogue 
($p=1.563 \pm 0.017 $~mas, $d= 640 \pm 7 $~pc, \citealt{GAIAcoll2016, GAIAcoll2018})
with the correction suggested by \citet{StassunTorres2018} of $-82 \pm 33\ \mu$as,
plus the photometry from the Two Micron All Sky Survey (2MASS, \citealt{Cutri2003,Skrutskie2006}).
For stellar models we used both MESA Isochrones \& Stellar Tracks (MIST, \citealt{Dotter2016,Choi2016,Paxton2011})
and the Dartmouth Stellar Evolution Database \citep{Dotter2008}.
We performed the analysis on the photospheric parameters obtained with HARPS-N as well as the literature values obtained by H10,
\cite{Huber2014}, \cite{Petigura2017} and \cite{Wang2018}.
For all methods we assumed 
$\sigma_{T_\mathrm{eff}} = 75$~K, $\sigma_{\log g} = 0.10$, $\sigma_{[{\rm Fe}/{\rm H}]} = 0.05$
as a good estimate of the systematic errors,
when the provided errors were lower than these values.\par

From the median and standard deviation of all the posterior samplings we obtained
\mstar~$=1.022_{-0.029}^{+0.039}$~\msun{} and $ \rstar = 0.96 \pm 0.02\ \rsun $.
We derived the stellar density \rhostar~$ = 1.16_{-0.09}^{+0.08}$~\rhosun{}
directly from the posterior distributions of \mstar{} and \rstar{}.
Following \cite{Lovis2011}, from the individual HARPS-N exposures
we measured a \logRHK{} index of $-4.67 \pm 0.09$, 
consistent with the young age of the star ($2.0^{+2.0}_{-1.3}$ Gy)
derived from the fit of the isochrones (\eg , \citealt{Pace2013}).
The astrophysical parameters of the star are summarised in Table~\ref{table:stellar_parameters},
where the temperature, gravity and metallicity are those obtained from the posterior distributions
to take into account the constraint from \textit{Gaia} parallax.
The stellar density determined in this work agrees very well with the value derived
in \citet{Freudenthal2018} with the dynamical analysis of the light curve.\par

\begin{table}
  \caption{Astrophysical parameters of the star.}
  \label{table:stellar_parameters}
  \centering
  \begin{tabular}{l c c }
  \hline
  Parameter & Value & Unit  \\
  \hline
  \noalign{\smallskip}
  2MASS alias & J19021775+3824032 & \\
  $\alpha_{\rm J2000}$ & 19:02:17.76 & hms \\
  $\delta_{\rm J2000}$ & +38:24:03.2 & dms \\
  \rstar\  & $ 0.958 \pm 0.020 $ & \rsun \\
  \mstar\ & $ 1.022_{-0.039}^{+0.029} $ & \msun \\
  \rhostar & $ 1.16_{-0.09}^{+0.08}$ & \rhosun  \\
  $\log (L_\star / L_\odot )$ & $ -0.038_{-0.027}^{0.026} $ & - \\
  $T_{{\rm eff}}$ & $ 5774 \pm 60 $ & K \\
  $\log g$ & $ 4.49_{-0.03}^{0.02}$ &-  \\
  $[{\rm Fe}/{\rm H}]$ & $ 0.05 \pm 0.07$ & - \\
  $p^{(a)}$ & $1.643 \pm 0.037$ & mas\\ 
  distance & $ 614 \pm 13 $ & pc \\
  A$_V$  & $0.10_{-0.07}^{+0.10}$ & mag \\
  B-V & $0.70 \pm 0.13 $ & mag \\
  age  & $ 2.0_{-1.3}^{+2.0}$ & Gy \\
  \logRHK & $-4.67 \pm 0.09$ & - \\
  \hline
  \end{tabular}
  \smallskip
  \begin{flushleft}
      $^{(a)}$ \textit{Gaia} parallax corrected for systematic offset as suggested by \citet{StassunTorres2018}.
  \end{flushleft}
\end{table}

\section{Analysis of the Kepler data}
\label{sec:KeplerLC}

We downloaded the {\it Kepler}-9 data from the MAST\footnote{Mikulsky Archive for Space Telescope, data release 25 (DR25).}
covering the full 17 quarters of the mission in long (LC) and short (SC) cadence.
We used the Presearch Data Conditioning (PDC) fluxes instead of Simple Aperture Photometry (SAP),
because the handling of systematic trends and errors were out of the purpose of this work.
We created a normalised full light curve by dividing the PDC flux of each quarter by its median value.
We selected carefully the portion of the transits and the out-of-transit,
in particular when the light curve shows transit events of \textit{Kepler}-9b and
\textit{Kepler}-9c very close to each other.\par

For each light curve we fitted the following parameters: 
$\log_{10}\rho_\star$ where $\rho_\star$ is the stellar density in \kgtomc, 
$\sqrt{k}$ where $k$ is the ratio between the planetary and the stellar radii, 
$\sqrt{b}$ where $b$ is the impact parameter\footnote{We used the equation of $b$
from \citet{Winn2011_arXiv} and \citet{Kipping2010}
that take into account not-zero eccentrity ($e$) and the argument of pericentre ($\omega$).},
the central time of the transit ($T_0$),
a quadratic limb-darkening (LD) law with the parameters $q_1$ and $q_2$ introduced in \citet{Kipping2013},
a linear trend (with $a_0$ as the intercept and $a_1$ the angular coefficient),
and $\log_2\sigma_j$, where $\sigma_j$ is a jitter term to add in quadrature to the errors of the \textit{Kepler} light curve.
We used an asymmetric prior for the $\log_{10}\rho_\star$
based on the stellar mass and radius values determined in section~\ref{sec:stellar_parameters},
while we used a uniform prior for the other parameters (see Table~\ref{tab:priors});
we kept fixed the period ($P$), the eccentricity and argument of pericentre (values from B14 solution).\par

We ran a Differential Evolution algorithm \citep[][\texttt{pyDE}\footnote{We used the python
implementation \texttt{pyDE} available on \url{https://github.com/hpparvi/PyDE}}]{Storn1997}
and then a Bayesian analysis of each selected light curve around each transit
by using the affine-invariant ensemble sampler \citep{GoodmanWeare2010} for Markov chain Monte Carlo (MCMC)
implemented within the \texttt{emcee} package
\citep{ForemanMackey2013} and modelling each transit with \texttt{batman} \citep{Kreidberg2015}.
We took into account the long exposure time of the LC data oversampling\footnote{We used an oversampling with a sub-exposure time of 9 seconds.}
the transit model;
for consistency, we used the same oversampling also for the SC data.\par

\begin{table*}
  \caption{Table of the priors, boundaries, 
  and final solution of the fitted parameters for transit analysis and the derived parameters.}
  \label{tab:priors}
	\begin{tabular}{lcccc}
	\hline
	parameter & prior type & boundaries [min, max] & \textit{Kepler}-9b & \textit{Kepler}-9c\\
	\hline
    $\log_{10}\rho_\star$  & G & $[0., 6.]$ & $3.196_{-0.060}^{+0.046}$ & $3.188_{-0.067}^{+0.048}$\\
    $\sqrt{k}$             & U & $[0., 1.]$ & $0.2786_{-0.0030}^{+0.0033}$ & $0.2750_{-0.0027}^{+0.0033}$ \\
    $\sqrt{b}$             & U & $\left[0., \sqrt{2}\right]$ & $0.767_{-0.035}^{+0.042}$ & $0.861_{-0.022}^{+0.024}$ \\
    $T_0$                   & U & $[T_0^\textrm{guess} \pm 1.5 \times T_{14}, ]$ & - & -\\
    $q_1$                  & U & $[0., 1.]$ & $0.43_{-0.26}^{+0.17}$ & $0.48_{-0.36}^{+0.20}$ \\
    $q_2$                  & U & $[0., 1.]$ & $0.18_{-0.18}^{+0.20}$ & $0.18_{-0.18}^{+0.16}$ \\
    $a_0$                  & U & $[-100., 100.]$ & - & -\\
    $a_1$                  & U & $[-100., 100.]$ & - & -\\
    $\log_2\sigma_j$       & U & $[\log_2(<\sigma_i>\times10^{-4}), 0.]$ & - & -\\
    \textit{derived transit model} &  & \\
    $\rho_\star$~(\rhosun) & - & - & $1.12_{-0.16}^{+0.11}$       & $1.10_{-0.17}^{+0.11}$ \\
    $a/R_\star$            & - & - & $31.3_{-1.5}^{+1.1}$         & $49.8_{-2.6}^{+1.7}$ \\
    $a$~(au)               & - & - & $0.143_{-0.006}^{+0.007}$    & $0.227_{-0.008}^{+0.012}$ \\
    $k$                    & - & - & $0.0776_{-0.0015}^{+0.0019}$ & $0.0756_{-0.0014}^{0.0018}$ \\
    $R_\prm$~$(R_\oplus)$  & - & - & $8.29_{-0.43}^{+0.54}$       & $8.08_{-0.41}^{+0.54}$ \\
    $b$                    & - & - & $0.59_{-0.05}^{0.06}$        & $0.74_{-0.04}^{+0.04}$ \\
    $i$~$(\degr)$          & - & - & $88.9_{-0.2}^{+0.1}$         & $89.1_{-0.1}^{+0.1}$ \\
    $T_{14}$~(min)         & - & - & $254.6_{-3.5}^{+7.9}$        & $273.9_{-7.3}^{+7.3}$ \\
    $u_1$                  & - & - & $0.24_{-0.24}^{+0.23}$       & $0.26_{-0.26}^{+0.20}$ \\
    $u_2$                  & - & - & $0.41_{-0.41}^{0.11}$        & $0.43_{-0.43}^{+0.14}$ \\
    \textit{linear ephemeris} &  &  & & \\
    $T_{0}$~(BJD$_\textrm{TDB}$) This work & - & - & $2454977.512 \pm 0.065$ & $2454968.84 \pm 0.20$ \\
    $P_\textrm{ephem}$~(days) This work & - & - & $19.2460 \pm 0.0015$ & $38.9492 \pm 0.0093$ \\
    \hline
  \end{tabular}
%   \medskip
  \smallskip
  \begin{flushleft}
      Notes: Prior type U means Uniform, while G means Gaussian (it could be an asymmetric Gaussian). 
      The stellar density prior has been computed as asymmetric Gaussian from $M_\star$ and $R_\star$.
      $T_0^\textrm{guess}$ has been obtained from the selection of each transit light curve, 
      while the $T_{14}$ is the transit total duration \citep[eq. 30 in][]{Kipping2010}.
      Following \citet{Kipping2013} we checked that the values of $q_1$ and $q_2$ were mapped to physical values
      of the quadratic LD coefficients $u_1$ and $u_2$.
      The linear trend coefficients ($a_0$, $a_1$) are bounded to very high values
      just to prevent singular behaviour.
      The minimum value for the $\log_{2}\sigma_{j}$ has been computed taking into account
      the mean value of the photometric errors ($\sigma_{i}$) of the portion of the light curve.
      We fixed the period, eccentricity, and argument of pericentre of both planets
      at the values of the best fit of B14.
  \end{flushleft}
\end{table*}

We fit each transit portion for both planets b and c for the SC with
50 walkers for 25000 generations with \texttt{pyDE} and then with 50000 steps with \texttt{emcee}.
Then we discarded as burn-in the first 25000 steps after checking the convergence
of the chains with the Gelman-Rubin (GR) statistics \citep[][\^{R}$=1.01$]{Gelman1992}.
For each transit, we obtained a final posterior after applying a thinning factor of 100,
\ie , a pessimistic estimate of the auto-correlation time of the chains.
All the posteriors from different transit fits were merged to obtain the final posterior of the parameters
$\log_{10}\rho_\star$, $\sqrt{k}$, $\sqrt{b}$, $q_1$, and $q_2$.\par

For each transit and merged posterior we computed the high density interval 
(HDI\footnote{Also indicated as high density region or high probability region.})
at $68.27\%$ (equivalent to $1\sigma$ error) 
for the fitted parameters and other physical quantities of interest derived from them.
As parameter estimation of each transit we selected the parameter set that maximise the likelihood
(maximum likelihood estimation, MLE) within the HDI,
while we computed the median of the merged posterior distributions.
The values of $\sqrt{k}$, $q_1$, and $q_2$ extracted from the SC merged posteriors
have been used as priors for the LC analysis (same number of walkers, generations, steps, and burn-in).
When we did not use the priors from the SC we found that the mean transit duration
for LC was $10\sigma$ longer than the SC for both planets.\par

\begin{figure}
\includegraphics[width=\linewidth]{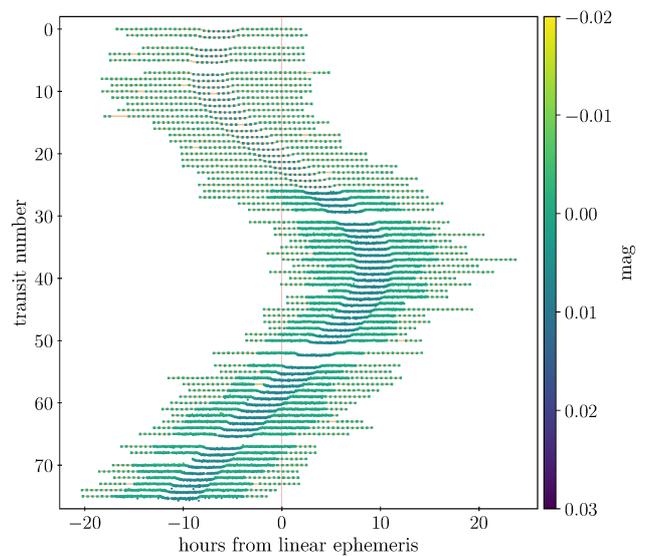}
\caption{River plot for \textit{Kepler}-9b:
the LC (and SC when available) of each transit has been plotted with colour code depending on the normalised magnitude 
(transit with darker colour, best-fit model in orange).
The light curves are sorted vertically by each epoch (or transit number with respect to a reference time)
as function of the phase (in hours) with respect to a linear ephemeris (vertical line at zero) 
showing the TTV effect.
}
\label{fig:riverK9b}
\end{figure}

\begin{figure}
\includegraphics[width=\linewidth]{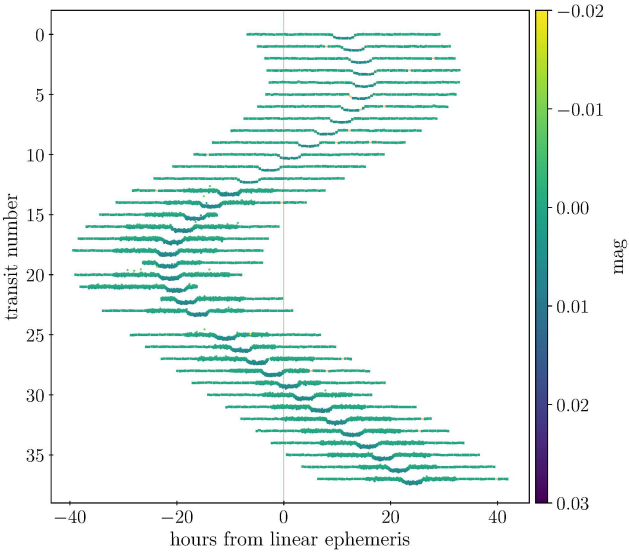}
\caption{Same as Fig.~\ref{fig:riverK9b}, but for \textit{Kepler}-9c.}
\label{fig:riverK9c}
\end{figure}

See Fig.~\ref{fig:riverK9b} and \ref{fig:riverK9c} for the river plots of planet b and c, respectively,
showing the data and the computed model of each transit in LC and SC and
the TTV effect with respect to a linear ephemeris.\par

As final parameters from the light curve analysis we decided
to use the mean between the SC and LC median values of the merged posteriors; 
we associated as lower uncertainty the lower value between the SC and LC
and the greater values for the upper uncertainty. 
The results are summarised in Table~\ref{tab:priors}.\par

\section{Orbital parameters from dynamical analysis}
\label{sec:dynamics}

We extract from the LC and SC analysis the central time, $T_0$,
and the total duration, $T_{14}$ 
\citep[defined as the difference between fourth and first contact time and computed with equation 30 in][]{Kipping2010},
of each transit; 
we kept the time and duration from SC when present, otherwise we used the analysis of the LC.
We assigned as symmetric errors of $T_0$s and $T_{14}$s the maximum between the lower and upper value of the HDI.
The measured central time and duration from each transit, in SC or LC mode, are available in Table~\ref{tab:tt_t41}.\par

For each planet, we used TRADES to fit the following parameters:
the mass of the planet in units of stellar mass, $M_\prm / M_\star $,
the planetary period $P_\prm$,
the eccentricity vector components $\sqrt{e_\prm}\cos(\omega_\prm)$ and $\sqrt{e_\prm}\sin(\omega_\prm)$,
where $e_\prm$ is the eccentricity and $\omega_\prm$ the argument of pericentre of the planet\footnote{We
stress the difference with $\omega_\star$ as defined by \cite{Eastman2013},
that is $\omega_\prm=\omega_\star+180\degr$.},
the inclination vector components  $i\cos(\Omega_\prm)$ and $i\sin(\Omega_\prm)$,
where $i$ is the orbital inclination and $\Omega_\prm$ the longitude of ascending node,
and the mean longitude of the planet $\lambda_\prm$,
defined as the sum of the  argument of pericentre, the longitude of the ascending node,
and the mean anomaly $\mathcal{M}_\prm$.
All the dynamical parameters are computed at the epoch of reference
BJD$_\textrm{TDB} = 2455088.212$.\par

We used uniform priors with broad but physically motivated boundaries:
planetary periods are constrained within two days of the value of the linear ephemeris,
and the mass of the planets, $M_\prm$, are bounded to less than $2\,M_\textrm{Jup}$.
We defined the reference coordinate system as described in 
\citet{Winn2011_arXiv}\footnote{Astrocentric reference system,with the plane $X-Y$ the sky-plane
and $Z$-axis pointing towards the observer, $X$-axis is aligned with line of nodes
and $\Omega$ fixed to $180\degr$.}, therefore \textit{Kepler}-9b has fixed $\Omega_\textrm{b}$ to $180\degr$,
effectively reducing the inclination vector of planet b to $i_\textrm{b}$.
All other parameters have been fixed to the values determined in section~\ref{sec:KeplerLC}, 
\eg, stellar and planetary radius.\par

\begin{figure*}
\includegraphics[width=\linewidth]{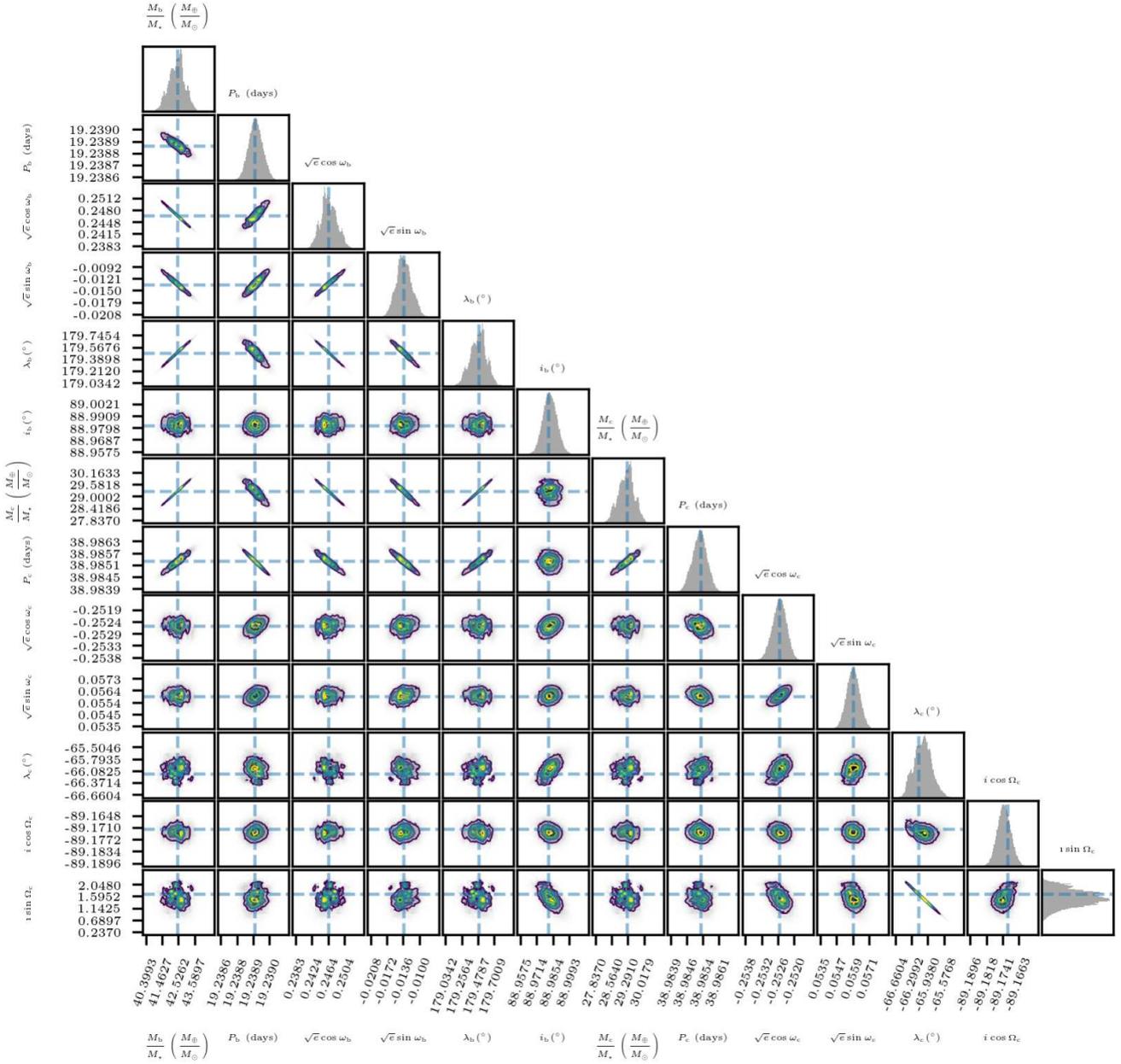}
\caption{Correlation plot of the fitted parameters.
The MLE solution has been shown as a dashed blue line.}
\label{fig:corrfit}
\end{figure*}

\begin{figure*}
\includegraphics[width=\linewidth]{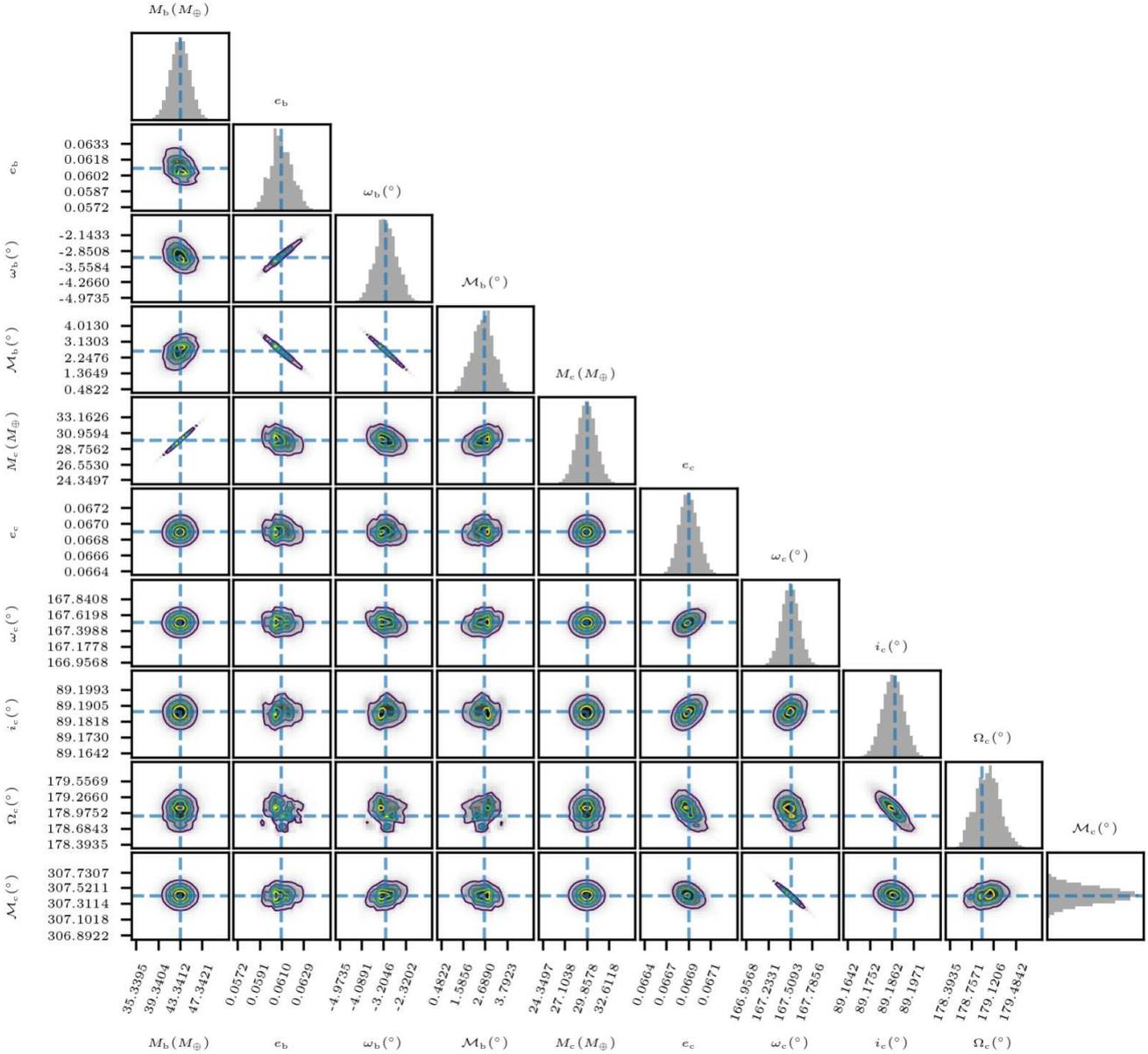}
\caption{Same as in Fig.~\ref{fig:corrfit}, but for derived parameters.}
\label{fig:corrder}
\end{figure*}

\begin{figure}
\includegraphics[width=\linewidth]{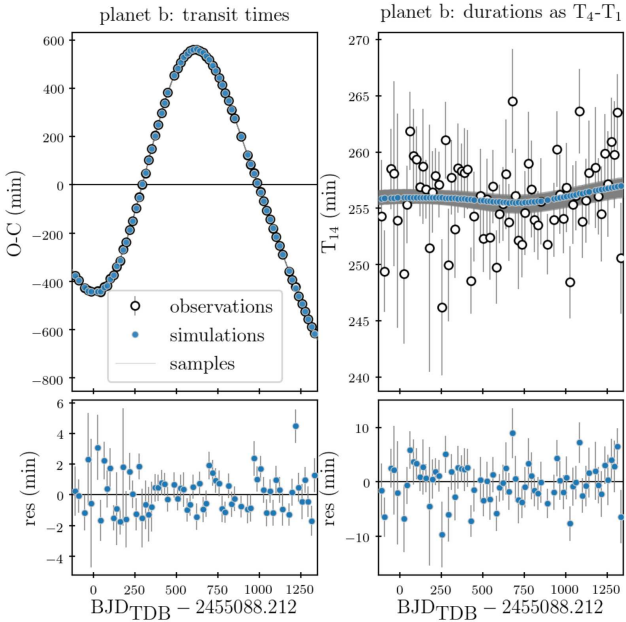}
\caption{O-C (upper-left panel) where O and C mean observed and calculated transit times, respectively
and duration (upper-right panel) plots with residuals (lower panels) for the best fit solution for \textit{Kepler}-9b.
Grey lines represent 1000 realisations obtained by randomly selecting sets of parameters from the posterior distributions.
}
\label{fig:K9b_oc}
\end{figure}

\begin{figure}
\includegraphics[width=\linewidth]{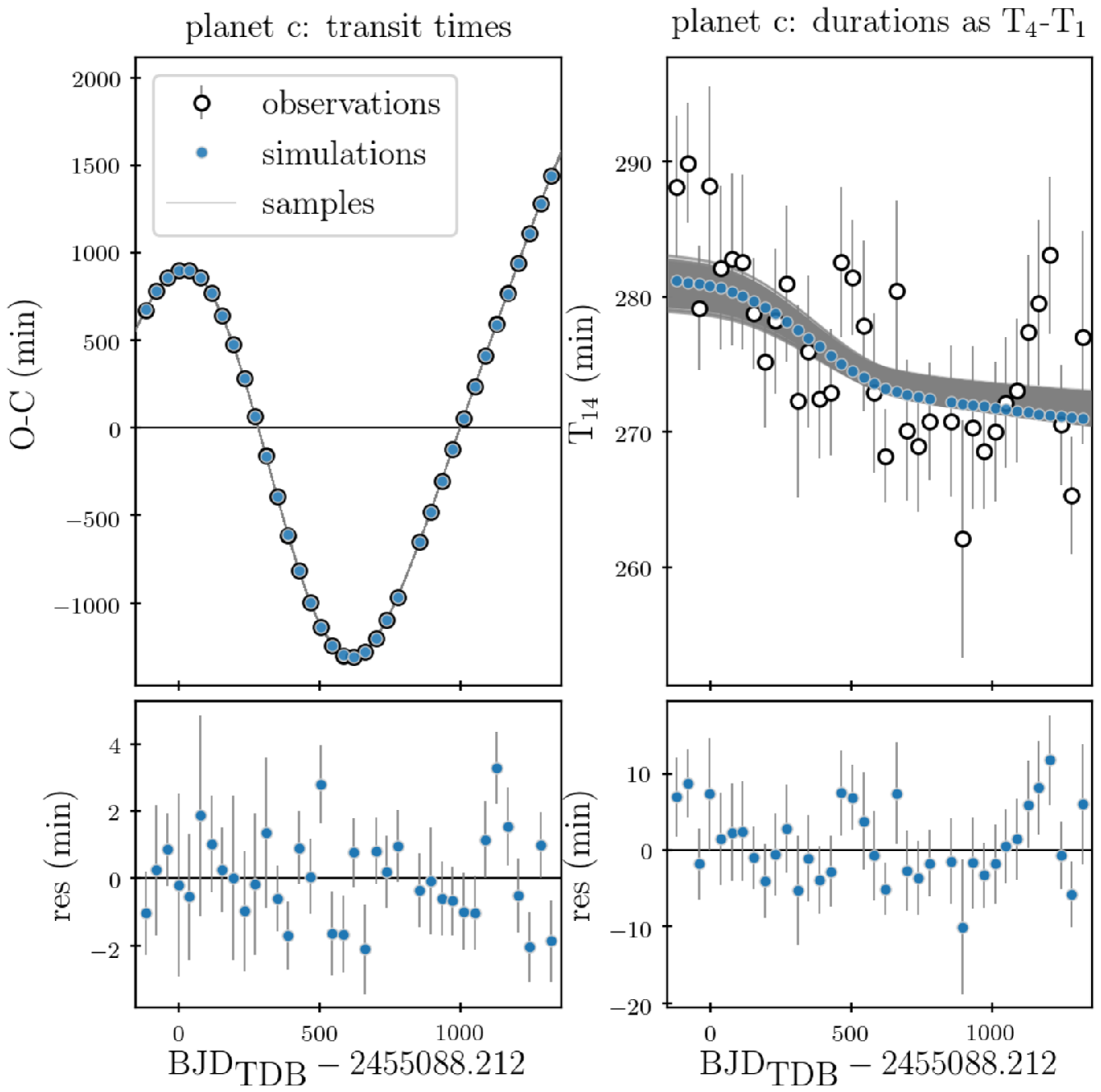}
\caption{Same as Fig.~\ref{fig:K9b_oc}, but for \textit{Kepler}-9c.}
\label{fig:K9c_oc}
\end{figure}

\begin{figure}
\includegraphics[width=\linewidth]{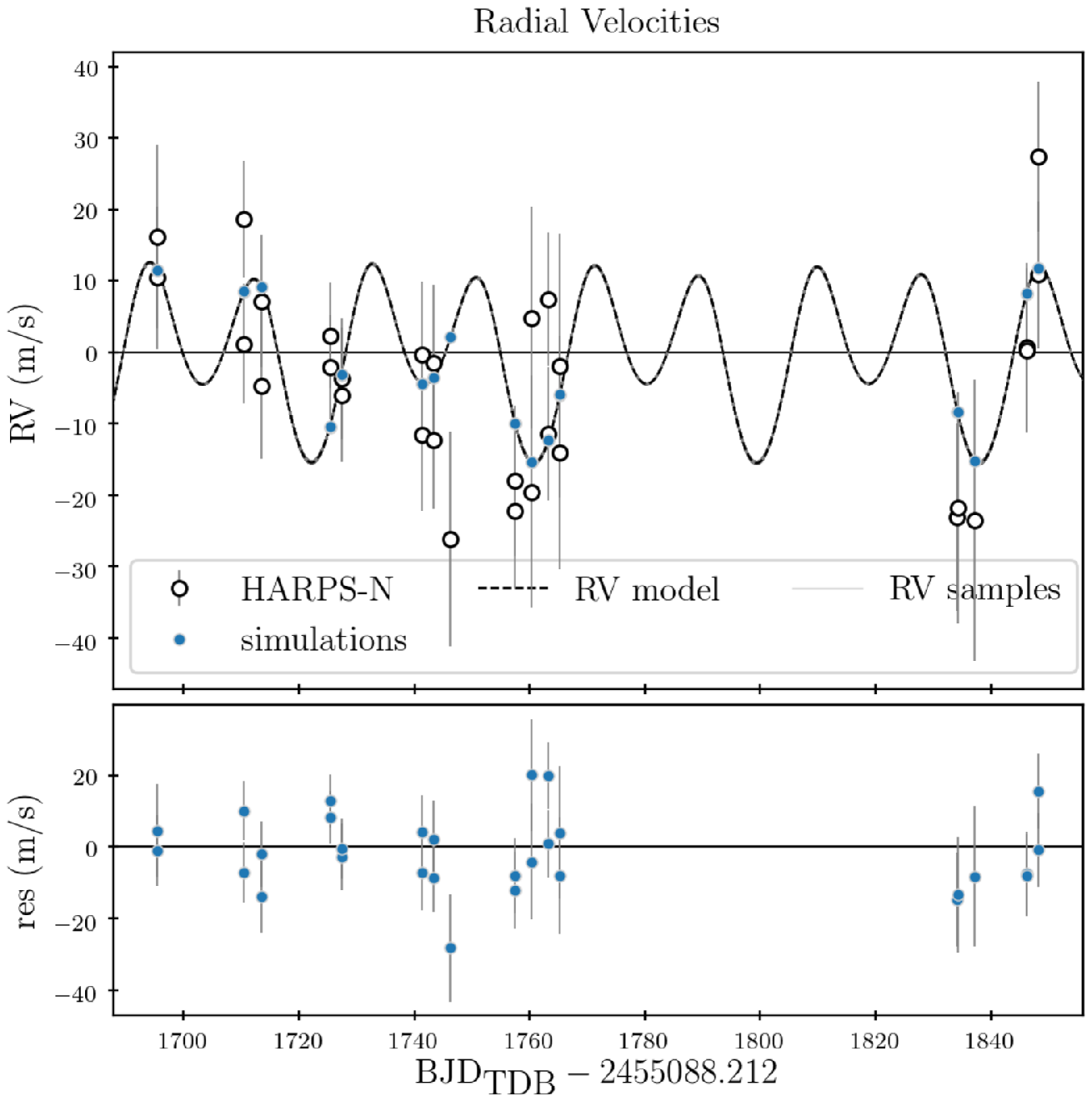}
\caption{\textit{Kepler}-9 RV plot (observed RVs as black-open circles) for best fit solution (blue-filled circles). 
The RV model of the best-fit solution covering the full integration time has been plotted as a black-dashed line.
We plotted 1000 realisations (as grey lines) of the model from the posterior distribution,
but they all lie too close to the best-fit RV model to be distinguishable from it.
}
\label{fig:K9_rv}
\end{figure}

We used an updated version of TRADES that allows us to fit $T_0$s and $T_{14}$s
simultaneously during the planetary orbit integration and
performs a Bayesian analysis with the \texttt{emcee} package;
we used the same form of the loglikelihood, $\ln{\mathcal{L}}$, introduced in \citet{Malavolta2017}.
Although {\tt TRADES} can also fit the observed RVs as well,
we did not include the HARPS-N data at this stage of the analysis because
we wanted to use those observations as an independent check of the TTV-derived solution.
We ran a simulation with 200 walkers, 200000 steps, and we discarded the first
50000 steps as burn-in\footnote{We visually checked the trace plot and we found that
a few walkers reached the convergence just before 50000 steps.}.
The initial walkers have been generated from a neighbourhood of the solution from B14.
As in the previous section we checked for GR statistics and
we obtained the final posterior distribution after the chains had been thinned with a factor of 100;
the correlation plots of the posterior distributions of the fitted and
derived parameters are shown in Fig~\ref{fig:corrfit} and \ref{fig:corrder}, respectively.
We computed the final best-fit parameter set as the MLE (and HDI) of the posterior distribution;
see the summary of the best-fit parameters in Table~\ref{tab:sumtable} and
how they reproduce the observed $T_0$ and $T_{14}$ data for \textit{Kepler}-9b and \textit{Kepler}-9c in 
Fig.~\ref{fig:K9b_oc} and \ref{fig:K9c_oc}, respectively. 
Figure~\ref{fig:K9_rv} shows how the new solution with the newly derived parameters fits the RV observations.
\par

The final solution confirmed the masses and the orbital parameters found in B14.\par

We also tested a parameter search fixing the inclination to the value from the LC analysis
and the longitude of node to $180\degr$, for both planets.
In this case the best-fit model found compatible masses but it did not reproduce the trend of the $T_{14}$ data
and the final Bayesian Information Criteria was higher.\par

\begin{table*}
  \caption{Table summarising the dynamical fit solution.
  Parameter values as the maximum likelihood estimation (MLE) and HDI at $68.27\%$ equivalent.
  Dynamical parameters are computed at the epoch of reference BJD$_\textrm{TDB}\, 2455088.212$.
  }
  \label{tab:sumtable}
  \begin{tabular}{lcc}
    \hline
    parameter & \textit{Kepler}-9b & \textit{Kepler}-9c\\
    \hline
    \textit{fitted dynamical model} &  & \\
    $M_\prm/M_\star$            & $0.000128_{-0.000002}^{+0.000001}$  & $0.000088_{-0.000001}^{+0.000001}$ \\
    $P_\prm$~(day)              & $19.23891_{-0.00006}^{+0.00006}$    & $38.9853_{-0.0003}^{+0.0003}$ \\
    $\sqrt{e}\cos(\omega_\prm)$ & $0.24651_{-0.0027}^{+0.0021}$       & $-0.2526_{-0.0003}^{+0.0003}$ \\
    $\sqrt{e}\sin(\omega_\prm)$ & $-0.014_{-0.002}^{+0.002}$          & $0.0559_{-0.0005}^{+0.0005}$ \\
    $i$~$(\degr)$               & $88.982_{-0.005}^{+0.007}$          & -- \\
    $i\cos(\Omega_\prm)$        & --                                  & $-89.172_{-0.005}^{+0.002}$ \\
    $i\sin(\Omega_\prm)$        & --                                  & $1.7_{-0.5}^{+0.2}$ \\
    $\lambda_\prm$~$(\degr)^\textrm{(a)}$  & $179.49_{-0.11}^{+0.15}$ & $293.9_{-0.1}^{+0.3}$ \\ % \lambda_c = -66
    \textit{derived dynamical model}  &  & \\
    $M_\prm$~$(M_\oplus)$         & $43.4_{-2.0}^{+1.6}$              & $29.9_{-1.3}^{+1.1}$ \\
    $\rho_\prm$~(g/cm$^{3}$)      & $0.42_{-0.09}^{+0.06}$            & $0.31_{-0.06}^{+0.05}$ \\
    $e_\prm$                      & $0.0609_{-0.0013}^{+0.0010}$      & $0.06691_{-0.00012}^{+0.00010}$ \\
    $\omega_\prm$~$(\degr)$       & $357.0_{-0.4}^{+0.5}$             & $167.5_{-0.1}^{+0.1}$ \\
    $\mathcal{M}_\prm$~$(\degr)$  & $2.6_{-0.6}^{+0.5}$               & $307.4_{-0.1}^{+0.1}$ \\ % mA_c = -52.58
    $i$~$(\degr)$                 & --                                & $89.188_{-0.006}^{+0.005}$ \\ 
    $\Omega_\prm$~$(\degr)$       & $180.$ (fixed)                    & $179.0_{-0.1}^{+0.3}$ \\
    \textit{dynamical model} $\chi^2_\textrm{r}$ (dof$=230$)          & 1.16 & \\
    \hline
    \multicolumn{3}{l}{$^\textrm{(a)}$ $\lambda_\prm$ is the mean longitude of the planet,
    defined as $\lambda_\prm = \Omega_\prm + \omega_\prm + \mathcal{M}_\prm$.}\\
  \end{tabular}
\end{table*}

\section{Comparison of literature solutions with RVs}

\begin{figure}
\includegraphics[width=\linewidth]{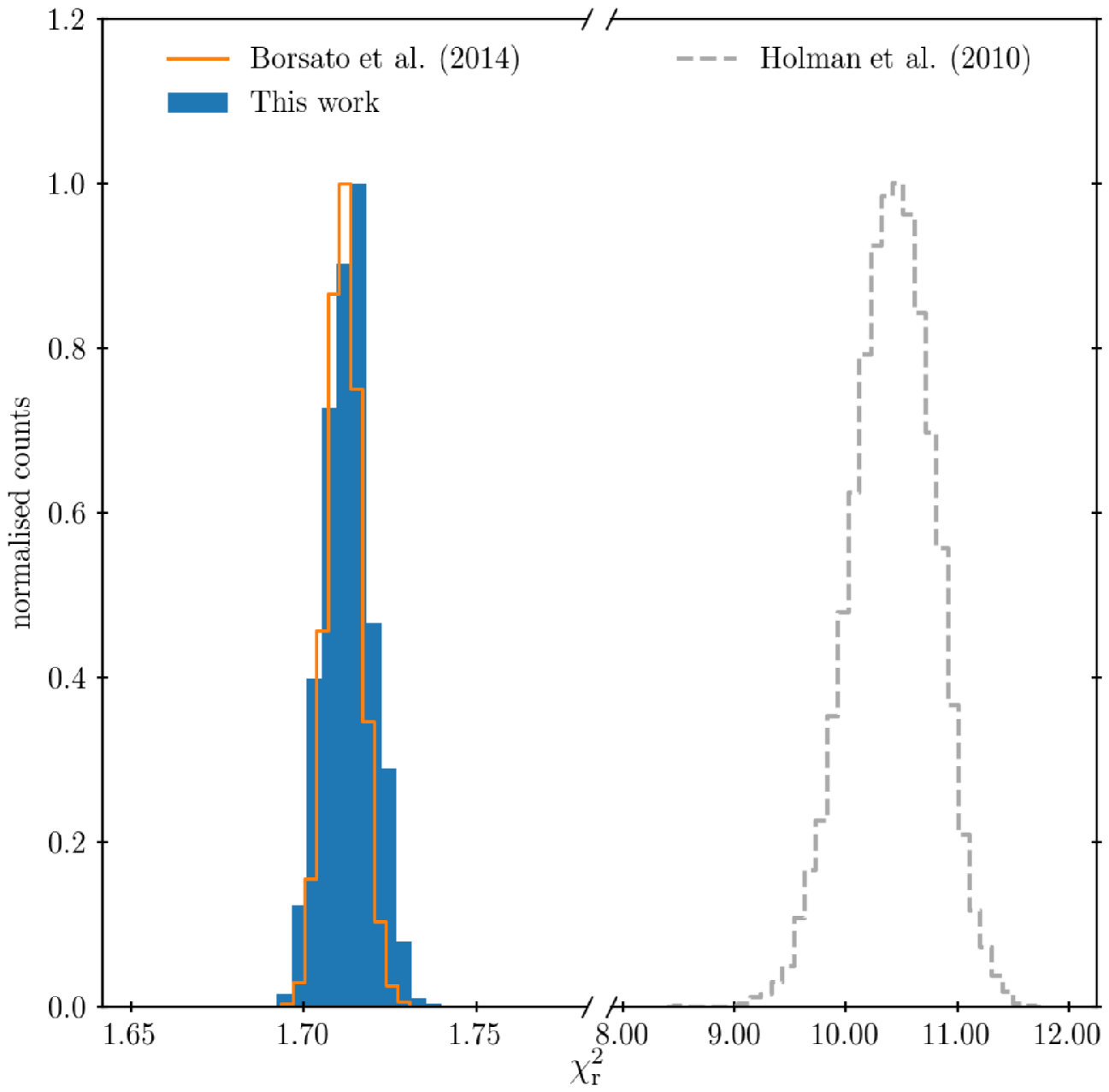}
\caption{Histograms of the $\chi^{2}_\textrm{r}$ from the RV of the 10000 simulations
for the three different solutions: H10 from \citep{Holman2010}, B14 from \citep{Borsato2014},
and from the posterior distribution of this work.
}
\label{fig:K9_vs}
\end{figure}

We propagated the solution of H10 and B14 to the epochs of our RVs
using the dynamical integrator embedded in TRADES \citep{Borsato2014}.
For consistency, we used the stellar masses of the two respective papers.
For each solution we generated 10000 sets by varying each parameter of
a quantity randomly extracted from a Gaussian distribution with 
variance equal to the errors listed in Table~\ref{tab:sumtable}.
We then computed the $\chi^2_\textrm{r}$ of the RVs for each set of orbital parameters.
In both cases we have only one free parameter, \ie , the systemic RV of the target, $\gamma$,
since all the other parameters have been determined independently of our RV dataset.
We can therefore use the $\chi^2_\textrm{r}$ to select which solution better reproduces our measurements.
We select 10000 sets of parameters from the posterior distribution computed in 
Sec.~\ref{sec:dynamics} and compute the $\chi^{2}_\textrm{r}$.
The distributions of the samples from B14 and this paper overlap each other and
they are centred at $\chi^2_\textrm{r}=1.71$,
while the H10 distribution has a higher $\chi^2_\textrm{r}$ of about 10.5 (see Fig.~\ref{fig:K9_vs}),
\ie , the solutions obtained using the $T_0$ alone
compare well with those from the HARPS-N RVs,
and do not depend on the exact number of transits involved in the analysis.\par

In our analysis we did not include the six Keck/HIRES observations gathered by H10,
since their apparent discrepancy with the TTV solutions presented by B14 and
following works was what motivated us obtaining new RVs with HARPS-N. 
While our RV dataset is well described by the TTV solutions  presented in B14 and in this work,
we still do not have an explanation for the inconsistency with the previous RVs from H10.
We note, however, that it is not the first time that a few, sparse Keck/HIRES RVs are 
in disagreement with a larger RV dataset obtained with HARPS-N
\citep[\eg,][]{Sozzetti2015}.
Given the observational strategy we employed, the large number of RVs,
and the inclusion of possible source of contamination from the Moon,
we believe that our dataset is more suitable to confirm or disprove
the possible disagreement between TTV and RVs methods.\par

\section{Discussion and Conclusions}

The apparent disagreement in the distribution of planets in the mass-radius diagram
between masses obtained from TTV modelling and masses derived from RVs
has been a long-standing problem in the exoplanet community
(see \citealt{Malavolta2017} for a review of the most interesting cases).
The \textit{Kepler}-9 system is the first system characterised with TTV,
and the already low densities of the planets 
as determined in the discovery paper by \citet{Holman2010}
have been further decreased in subsequent analysis by \citet{Borsato2014}.
Consequently, we selected this system as a proxy to compare RV and TTV masses measurements.\par

A differential comparison between the predicted RVs from H10 and B14 allowed us
to set up the best observational strategy to be carried out with HARPS-N 
in order to independently confirm the true nature of the \textit{Kepler}-9 planets
and to overcome the fact that the faintness of the star would make a 5-$\sigma$ detection unfeasible.
Additionally, we performed an independent analysis of the full \textit{Kepler} light-curve 
in order to compare the observed RVs with the prediction of
the most precise orbital model that could be obtained from TTVs.
In our analysis we also included refined stellar parameters
that take into account the precise parallax measurement provided by \textit{Gaia} DR2,
and both literature and newly derived photospheric parameters for the star.\par

\begin{figure}
\includegraphics[width=\linewidth]{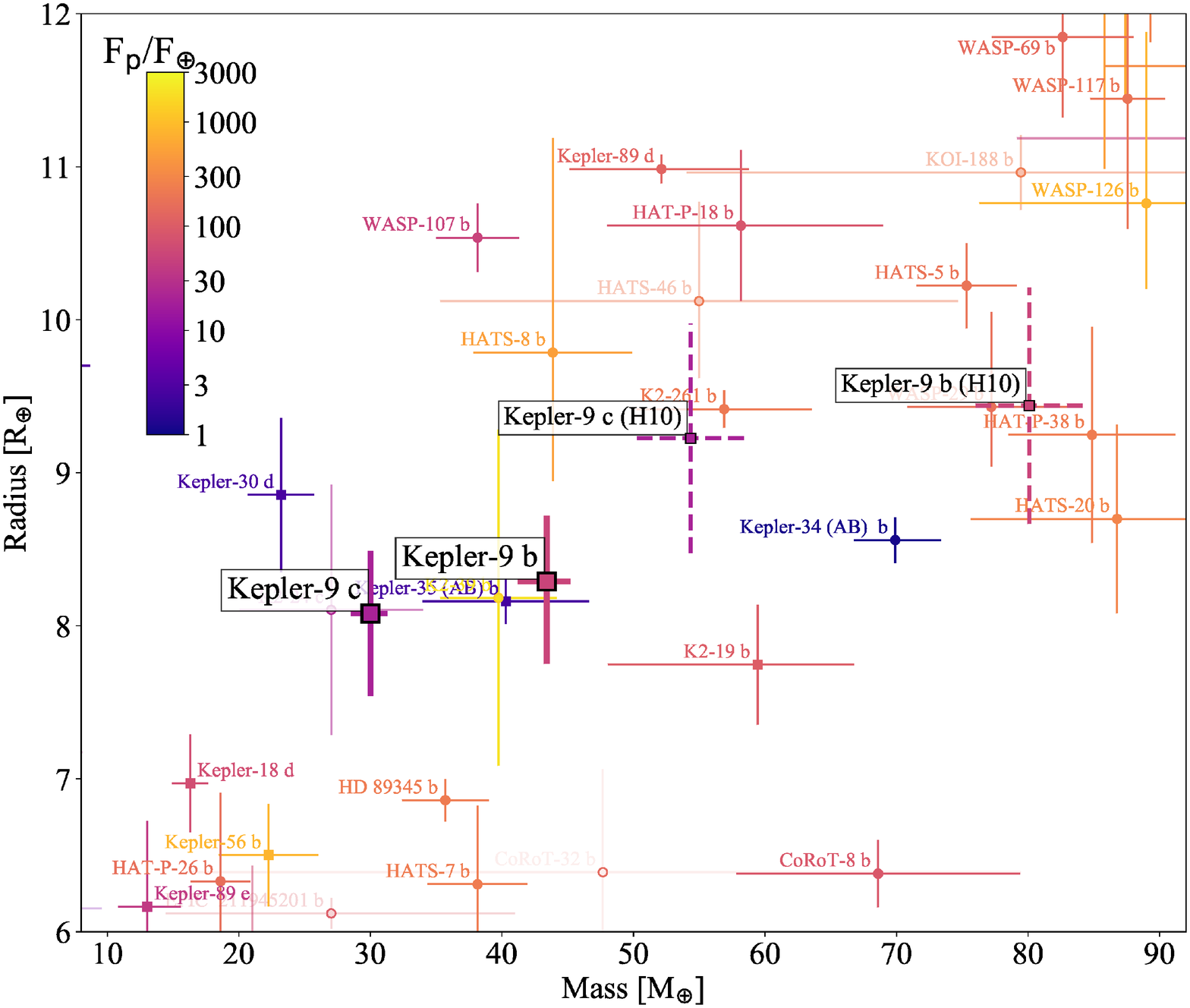}
\caption{Mass-Radius plot for \textit{Kepler}-9b and c.
Results from H10 are included for comparisons.
Planets with mass determination with uncertainties greater
than 20\% are shaded accordingly.
Planets are color-coded according to their incident flux.
Data obtained from \url{exoplanet.eu} on October 2018.
}
\label{fig:K9_MR}
\end{figure}

The fit of the whole \textit{Kepler} data-set confirmed the masses, 
and hence the densities (see Table~\ref{tab:sumtable}),
originally found by B14. Our mass values are also well consistent
with those in the recent work by \citet{Freudenthal2018},
meaning that with a dynamical model we have been able to obtain masses 
with the same precision as the more computationally expensive photodynamical model,
even when using space-borne photometry alone. Our analysis places \mbox{\textit{Kepler}-9b} and c 
in the mass-radius region of super-Neptunes / inflated sub-Saturns together with
other planets recently discovered that also have precise RV-derived masses,
such as WASP-139b \citep{Hellier2017}, K2-24c \citep{Dai2016}, K2-39b \citep{VanEylen2016} 
and HATS-8b \citep{Bayliss2015}, as can be seen in Figure~\ref{fig:K9_MR}.\par

The observed HARPS-N RVs noticeably agree with the model of this paper (and B14)
with a $\chi^{2}_\textrm{r}$ of 1.71,
but are not consistent with the solution from H10 
($\chi^{2}_\textrm{r} \sim 10.5$).
The high value of the masses found by H10 is ascribable to
the very short baseline of the photometric data
and due to a possible underestimation of the HIRES RV uncertainties.\par

The \textit{Kepler} mission has shown the power of the TTV method
to determine the planetary nature of candidates in multiple systems and
to characterise the orbital and physical parameters of the exoplanets.
The lack of bright stars hosting multiple-planet systems showing TTV
in the \textit{Kepler} field did not allow for high precision RV observation
in order to confirm or rule out the mass determination discrepancy.
With the advent of the TESS \citep{Ricker2014} and PLATO \citep{Rauer2014} missions 
we will be able to measure the masses of many planets around bright stars
with both the TTV and the RV methods allowing us to further investigate
the level of consistency in planet parameters between the two methods,
or the presence of any bias in the solutions coming from the two techniques.
\par

\section*{Acknowledgments}
Based on observations made with the Italian Telescopio Nazionale Galileo (TNG)
operated on the island of La Palma by the Fundacion Galileo Galilei of the
INAF (Istituto Nazionale di Astrofisica) at the Spanish Observatorio
del Roque de los Muchachos of the Instituto de Astrofisica de Canarias.
The HARPS-N project has been funded by the Prodex Program of the Swiss Space Office (SSO),
the Harvard University Origins of Life Initiative (HUOLI),
the Scottish Universities Physics Alliance (SUPA),
the University of Geneva, the Smithsonian Astrophysical Observatory (SAO),
and the Italian National Astrophysical Institute (INAF), the University of St Andrews,
Queen's University Belfast, and the University of Edinburgh.
The research leading to these results received funding from 
the European Union Seventh Framework Programme (FP7/2007- 2013) under grant agreement number 313014 (ETAEARTH)
and support from Italian Space Agency (ASI) regulated by Accordo ASI-INAF n. 2013-016-R.0 
del 9 luglio 2013 e integrazione del 9 luglio 2015.
This work has made use of data from the European Space Agency (ESA) mission
\textit{Gaia} (\url{https://www.cosmos.esa.int/gaia}), processed by the \textit{Gaia}
Data Processing and Analysis Consortium (DPAC,
\url{https://www.cosmos.esa.int/web/gaia/dpac/consortium}). Funding for the DPAC
has been provided by national institutions, in particular the institutions
participating in the \textit{Gaia} Multilateral Agreement.
LM acknowledge the support by INAF/Frontiera through the 
"Progetti Premiali" funding scheme of the Italian Ministry of Education, 
University, and Research.
ACC acknowledges support from the Science \&\ Technology Facilities Council (STFC)
consolidated grant number ST/R000824/1.
Some of this work has been carried out within the framework of the NCCR PlanetS,
supported by the Swiss National Science Foundation.
X. D. is grateful to the Branco-Weiss Fellowship-Society in Science for its financial support.
CAW acknowledges support from the STFC grant ST/P000312/1.
DWL acknowledges partial support from the Kepler mission under
NASA Cooperative Agreement NNX13AB58A with
the Smithsonian Astrophysical Observatory.
This material is based upon work supported by 
the National Aeronautics and Space Administration under grants
No. NNX15AC90G and NNX17AB59G issued through the Exoplanets Research Program.
This research has made use of the Exoplanet Follow-up Observation Program, 
NASA's Astrophysics Data System and
the NASA Exoplanet Archive, which are operated by
the California Institute of Technology, under contract with
the National Aeronautics and Space Administration under the Exoplanet Exploration Program.
Some of the data presented in this paper were obtained from
the Mikulski Archive for Space Telescopes (MAST).
STScI is operated by the Association of Universities
for Research in Astronomy, Inc., under NASA contract NAS5-26555.
Support for MAST for non-HST data is provided by the NASA Office
of Space Science via grant NNX13AC07G and by other grants and contracts.
\par

\bibliographystyle{mn2e}
\bibliography{Bibliography}

\begin{thebibliography}{63}
\expandafter\ifx\csname natexlab\endcsname\relax\def\natexlab#1{#1}\fi

\bibitem[{{Agol} {et~al}\mbox{.}(2005){Agol}, {Steffen}, {Sari}, \&
  {Clarkson}}]{Agol2005}
{Agol} E., {Steffen} J., {Sari} R., {Clarkson} W., 2005, \mnras, 359, 567

\bibitem[{{Barros} {et~al}\mbox{.}(2015){Barros}, {Almenara}, {Demangeon},
  {Tsantaki}, {Santerne}, {Armstrong}, {Barrado}, {Brown}, {Deleuil},
  {Lillo-Box}, {Osborn}, {Pollacco}, {Abe}, {Andre}, {Bendjoya}, {Boisse},
  {Bonomo}, {Bouchy}, {Bruno}, {Cerda}, {Courcol}, {D{\'{\i}}az},
  {H{\'e}brard}, {Kirk}, {Lachuri{\'e}}, {Lam}, {Martinez}, {McCormac},
  {Moutou}, {Rajpurohit}, {Rivet}, {Spake}, {Suarez}, {Toublanc}, \&
  {Walker}}]{Barros2015}
{Barros} S.~C.~C. {et~al.}, 2015, \mnras, 454, 4267

\bibitem[{{Bayliss} {et~al}\mbox{.}(2015){Bayliss}, {Hartman}, {Bakos},
  {Penev}, {Zhou}, {Brahm}, {Rabus}, {Jord{\'a}n}, {Mancini}, {de Val-Borro},
  {Bhatti}, {Espinoza}, {Csubry}, {Howard}, {Fulton}, {Buchhave}, {Henning},
  {Schmidt}, {Ciceri}, {Noyes}, {Isaacson}, {Marcy}, {Suc}, {L{\'a}z{\'a}r},
  {Papp}, \& {S{\'a}ri}}]{Bayliss2015}
{Bayliss} D. {et~al.}, 2015, \aj, 150, 49

\bibitem[{{Becker} {et~al}\mbox{.}(2015){Becker}, {Vanderburg}, {Adams},
  {Rappaport}, \& {Schwengeler}}]{Becker2015}
{Becker} J.~C., {Vanderburg} A., {Adams} F.~C., {Rappaport} S.~A.,
  {Schwengeler} H.~M., 2015, \apjl, 812, L18

\bibitem[{{Borsato} {et~al}\mbox{.}(2014){Borsato}, {Marzari}, {Nascimbeni},
  {Piotto}, {Granata}, {Bedin}, \& {Malavolta}}]{Borsato2014}
{Borsato} L., {Marzari} F., {Nascimbeni} V., {Piotto} G., {Granata} V., {Bedin}
  L.~R., {Malavolta} L., 2014, \aap, 571, A38

\bibitem[{{Borucki} {et~al}\mbox{.}(2011){Borucki}, {Koch}, {Basri}, {Batalha},
  {Brown}, {Bryson}, {Caldwell}, {Christensen-Dalsgaard}, {Cochran}, {DeVore},
  {Dunham}, {Gautier}, {Geary}, {Gilliland}, {Gould}, {Howell}, {Jenkins},
  {Latham}, {Lissauer}, {Marcy}, {Rowe}, {Sasselov}, {Boss}, {Charbonneau},
  {Ciardi}, {Doyle}, {Dupree}, {Ford}, {Fortney}, {Holman}, {Seager},
  {Steffen}, {Tarter}, {Welsh}, {Allen}, {Buchhave}, {Christiansen}, {Clarke},
  {Das}, {D{\'e}sert}, {Endl}, {Fabrycky}, {Fressin}, {Haas}, {Horch},
  {Howard}, {Isaacson}, {Kjeldsen}, {Kolodziejczak}, {Kulesa}, {Li}, {Lucas},
  {Machalek}, {McCarthy}, {MacQueen}, {Meibom}, {Miquel}, {Prsa}, {Quinn},
  {Quintana}, {Ragozzine}, {Sherry}, {Shporer}, {Tenenbaum}, {Torres},
  {Twicken}, {Van Cleve}, {Walkowicz}, {Witteborn}, \& {Still}}]{Borucki2011}
{Borucki} W.~J. {et~al.}, 2011, \apj, 736, 19

\bibitem[{{Buchhave} {et~al}\mbox{.}(2014){Buchhave}, {Bizzarro}, {Latham},
  {Sasselov}, {Cochran}, {Endl}, {Isaacson}, {Juncher}, \&
  {Marcy}}]{Buchhave2014}
{Buchhave} L.~A. {et~al.}, 2014, \nat, 509, 593

\bibitem[{{Buchhave} {et~al}\mbox{.}(2012){Buchhave}, {Latham}, {Johansen},
  {Bizzarro}, {Torres}, {Rowe}, {Batalha}, {Borucki}, {Brugamyer}, {Caldwell},
  {Bryson}, {Ciardi}, {Cochran}, {Endl}, {Esquerdo}, {Ford}, {Geary},
  {Gilliland}, {Hansen}, {Isaacson}, {Laird}, {Lucas}, {Marcy}, {Morse},
  {Robertson}, {Shporer}, {Stefanik}, {Still}, \& {Quinn}}]{Buchhave2012}
{Buchhave} L.~A. {et~al.}, 2012, \nat, 486, 375

\bibitem[{{Choi} {et~al}\mbox{.}(2016){Choi}, {Dotter}, {Conroy}, {Cantiello},
  {Paxton}, \& {Johnson}}]{Choi2016}
{Choi} J., {Dotter} A., {Conroy} C., {Cantiello} M., {Paxton} B., {Johnson}
  B.~D., 2016, \apj, 823, 102

\bibitem[{{Cochran} {et~al}\mbox{.}(2011){Cochran}, {Fabrycky}, {Torres},
  {Fressin}, {D{\'e}sert}, {Ragozzine}, {Sasselov}, {Fortney}, {Rowe},
  {Brugamyer}, {Bryson}, {Carter}, {Ciardi}, {Howell}, {Steffen}, {Borucki},
  {Koch}, {Winn}, {Welsh}, {Uddin}, {Tenenbaum}, {Still}, {Seager}, {Quinn},
  {Mullally}, {Miller}, {Marcy}, {MacQueen}, {Lucas}, {Lissauer}, {Latham},
  {Knutson}, {Kinemuchi}, {Johnson}, {Jenkins}, {Isaacson}, {Howard}, {Horch},
  {Holman}, {Henze}, {Haas}, {Gilliland}, {Gautier}, {Ford}, {Fischer},
  {Everett}, {Endl}, {Demory}, {Deming}, {Charbonneau}, {Caldwell}, {Buchhave},
  {Brown}, \& {Batalha}}]{Cochran2011}
{Cochran} W.~D. {et~al.}, 2011, \apjs, 197, 7

\bibitem[{{Cutri} {et~al}\mbox{.}(2003){Cutri}, {Skrutskie}, {van Dyk},
  {Beichman}, {Carpenter}, {Chester}, {Cambresy}, {Evans}, {Fowler}, {Gizis},
  {Howard}, {Huchra}, {Jarrett}, {Kopan}, {Kirkpatrick}, {Light}, {Marsh},
  {McCallon}, {Schneider}, {Stiening}, {Sykes}, {Weinberg}, {Wheaton},
  {Wheelock}, \& {Zacarias}}]{Cutri2003}
{Cutri} R.~M. {et~al.}, 2003, VizieR Online Data Catalog, 2246

\bibitem[{{Dai} {et~al}\mbox{.}(2016){Dai}, {Winn}, {Albrecht}, {Arriagada},
  {Bieryla}, {Butler}, {Crane}, {Hirano}, {Johnson}, {Kiilerich}, {Latham},
  {Narita}, {Nowak}, {Palle}, {Ribas}, {Rogers}, {Sanchis-Ojeda}, {Shectman},
  {Teske}, {Thompson}, {Van Eylen}, {Vanderburg}, {Wittenmyer}, \&
  {Yu}}]{Dai2016}
{Dai} F. {et~al.}, 2016, \apj, 823, 115

\bibitem[{{Dotter}(2016)}]{Dotter2016}
{Dotter} A., 2016, \apjs, 222, 8

\bibitem[{{Dotter} {et~al}\mbox{.}(2008){Dotter}, {Chaboyer}, {Jevremovi{\'c}},
  {Kostov}, {Baron}, \& {Ferguson}}]{Dotter2008}
{Dotter} A., {Chaboyer} B., {Jevremovi{\'c}} D., {Kostov} V., {Baron} E.,
  {Ferguson} J.~W., 2008, \apjs, 178, 89

\bibitem[{{Eastman}, {Gaudi} \& {Agol}(2013){Eastman}, {Gaudi}, \&
  {Agol}}]{Eastman2013}
{Eastman} J., {Gaudi} B.~S., {Agol} E., 2013, \pasp, 125, 83

\bibitem[{{Feroz} \& {Hobson}(2008)}]{Feroz2008}
{Feroz} F., {Hobson} M.~P., 2008, \mnras, 384, 449

\bibitem[{{Feroz}, {Hobson} \& {Bridges}(2009){Feroz}, {Hobson}, \&
  {Bridges}}]{Feroz2009}
{Feroz} F., {Hobson} M.~P., {Bridges} M., 2009, \mnras, 398, 1601

\bibitem[{{Feroz} {et~al}\mbox{.}(2013){Feroz}, {Hobson}, {Cameron}, \&
  {Pettitt}}]{Feroz2013}
{Feroz} F., {Hobson} M.~P., {Cameron} E., {Pettitt} A.~N., 2013, ArXiv e-prints

\bibitem[{{Foreman-Mackey} {et~al}\mbox{.}(2013){Foreman-Mackey}, {Hogg},
  {Lang}, \& {Goodman}}]{ForemanMackey2013}
{Foreman-Mackey} D., {Hogg} D.~W., {Lang} D., {Goodman} J., 2013, \pasp, 125,
  306

\bibitem[{{Freudenthal} {et~al}\mbox{.}(2018){Freudenthal}, {von Essen},
  {Dreizler}, {Wedemeyer}, {Agol}, {Morris}, {Becker}, {Mallonn}, {Hoyer},
  {Ofir}, {Tal- Or}, {Deeg}, {Herrero}, {Ribas}, {Khalafinejad},
  {Hern{\'a}ndez}, \& {Rodr{\'\i}guez S.}}]{Freudenthal2018}
{Freudenthal} J. {et~al.}, 2018, \aap, 618, A41

\bibitem[{{Gaia Collaboration} {et~al}\mbox{.}(2018){Gaia Collaboration},
  {Brown}, {Vallenari}, {Prusti}, {de Bruijne}, {Babusiaux}, \&
  {Bailer-Jones}}]{GAIAcoll2018}
{Gaia Collaboration}, {Brown} A.~G.~A., {Vallenari} A., {Prusti} T., {de
  Bruijne} J.~H.~J., {Babusiaux} C., {Bailer-Jones} C.~A.~L., 2018, ArXiv
  e-prints

\bibitem[{{Gaia Collaboration} {et~al}\mbox{.}(2016){Gaia Collaboration},
  {Prusti}, {de Bruijne}, {Brown}, {Vallenari}, {Babusiaux}, {Bailer-Jones},
  {Bastian}, {Biermann}, {Evans}, \& et~al.}]{GAIAcoll2016}
{Gaia Collaboration} {et~al.}, 2016, \aap, 595, A1

\bibitem[{{Gelman} \& {Rubin}(1992)}]{Gelman1992}
{Gelman} A., {Rubin} D.~B., 1992, Statistical Science, 7, 457

\bibitem[{{Goodman} \& {Weare}(2010)}]{GoodmanWeare2010}
{Goodman} J., {Weare} J., 2010, Communications in Applied Mathematics and
  Computational Science, Vol.~5, No.~1, p.~65-80, 2010, 5, 65

\bibitem[{{Hellier} {et~al}\mbox{.}(2017){Hellier}, {Anderson}, {Cameron},
  {Delrez}, {Gillon}, {Jehin}, {Lendl}, {Maxted}, {Neveu-VanMalle}, {Pepe},
  {Pollacco}, {Queloz}, {S{\'e}gransan}, {Smalley}, {Southworth}, {Triaud},
  {Udry}, {Wagg}, \& {West}}]{Hellier2017}
{Hellier} C. {et~al.}, 2017, \mnras, 465, 3693

\bibitem[{{Holman} {et~al}\mbox{.}(2010){Holman}, {Fabrycky}, {Ragozzine},
  {Ford}, {Steffen}, {Welsh}, {Lissauer}, {Latham}, {Marcy}, {Walkowicz},
  {Batalha}, {Jenkins}, {Rowe}, {Cochran}, {Fressin}, {Torres}, {Buchhave},
  {Sasselov}, {Borucki}, {Koch}, {Basri}, {Brown}, {Caldwell}, {Charbonneau},
  {Dunham}, {Gautier}, {Geary}, {Gilliland}, {Haas}, {Howell}, {Ciardi},
  {Endl}, {Fischer}, {F{\"u}r{\'e}sz}, {Hartman}, {Isaacson}, {Johnson},
  {MacQueen}, {Moorhead}, {Morehead}, \& {Orosz}}]{Holman2010}
{Holman} M.~J. {et~al.}, 2010, Science, 330, 51

\bibitem[{{Holman} \& {Murray}(2005)}]{HolmanMurray2005}
{Holman} M.~J., {Murray} N.~W., 2005, Science, 307, 1288

\bibitem[{{Huber} {et~al}\mbox{.}(2014){Huber}, {Silva Aguirre}, {Matthews},
  {Pinsonneault}, {Gaidos}, {Garc{\'{\i}}a}, {Hekker}, {Mathur}, {Mosser},
  {Torres}, {Bastien}, {Basu}, {Bedding}, {Chaplin}, {Demory}, {Fleming},
  {Guo}, {Mann}, {Rowe}, {Serenelli}, {Smith}, \& {Stello}}]{Huber2014}
{Huber} D. {et~al.}, 2014, \apjs, 211, 2

\bibitem[{{Jontof-Hutter} {et~al}\mbox{.}(2016){Jontof-Hutter}, {Ford}, {Rowe},
  {Lissauer}, {Fabrycky}, {Van Laerhoven}, {Agol}, {Deck}, {Holczer}, \&
  {Mazeh}}]{JontofHutter2016}
{Jontof-Hutter} D. {et~al.}, 2016, \apj, 820, 39

\bibitem[{{Kipping}(2010)}]{Kipping2010}
{Kipping} D.~M., 2010, \mnras, 407, 301

\bibitem[{{Kipping}(2013)}]{Kipping2013}
{Kipping} D.~M., 2013, \mnras, 435, 2152

\bibitem[{{Kreidberg}(2015)}]{Kreidberg2015}
{Kreidberg} L., 2015, \pasp, 127, 1161

\bibitem[{{Lissauer} {et~al}\mbox{.}(2011){Lissauer}, {Fabrycky}, {Ford},
  {Borucki}, {Fressin}, {Marcy}, {Orosz}, {Rowe}, {Torres}, {Welsh}, {Batalha},
  {Bryson}, {Buchhave}, {Caldwell}, {Carter}, {Charbonneau}, {Christiansen},
  {Cochran}, {Desert}, {Dunham}, {Fanelli}, {Fortney}, {Gautier}, {Geary},
  {Gilliland}, {Haas}, {Hall}, {Holman}, {Koch}, {Latham}, {Lopez},
  {McCauliff}, {Miller}, {Morehead}, {Quintana}, {Ragozzine}, {Sasselov},
  {Short}, \& {Steffen}}]{Lissauer2011}
{Lissauer} J.~J. {et~al.}, 2011, \nat, 470, 53

\bibitem[{{Lovis} {et~al}\mbox{.}(2011){Lovis}, {Dumusque}, {Santos}, {Bouchy},
  {Mayor}, {Pepe}, {Queloz}, {S{\'e}gransan}, \& {Udry}}]{Lovis2011}
{Lovis} C. {et~al.}, 2011, ArXiv e-prints

\bibitem[{{Malavolta} {et~al}\mbox{.}(2017{\natexlab{a}}){Malavolta},
  {Borsato}, {Granata}, {Piotto}, {Lopez}, {Vanderburg}, {Figueira}, {Mortier},
  {Nascimbeni}, {Affer}, {Bonomo}, {Bouchy}, {Buchhave}, {Charbonneau},
  {Collier Cameron}, {Cosentino}, {Dressing}, {Dumusque}, {Fiorenzano},
  {Harutyunyan}, {Haywood}, {Johnson}, {Latham}, {Lopez-Morales}, {Lovis},
  {Mayor}, {Micela}, {Molinari}, {Motalebi}, {Pepe}, {Phillips}, {Pollacco},
  {Queloz}, {Rice}, {Sasselov}, {S{\'e}gransan}, {Sozzetti}, {Udry}, \&
  {Watson}}]{Malavolta2017}
{Malavolta} L. {et~al.}, 2017{\natexlab{a}}, \aj, 153, 224

\bibitem[{{Malavolta} {et~al}\mbox{.}(2017{\natexlab{b}}){Malavolta}, {Lovis},
  {Pepe}, {Sneden}, \& {Udry}}]{Malavolta2017b}
{Malavolta} L., {Lovis} C., {Pepe} F., {Sneden} C., {Udry} S.,
  2017{\natexlab{b}}, \mnras, 469, 3965

\bibitem[{{Malavolta} {et~al}\mbox{.}(2018){Malavolta}, {Mayo}, {Louden},
  {Rajpaul}, {Bonomo}, {Buchhave}, {Kreidberg}, {Kristiansen}, {Lopez-
  Morales}, {Mortier}, {Vanderburg}, {Coffinet}, {Ehrenreich}, {Lovis},
  {Bouchy}, {Charbonneau}, {Ciardi}, {Collier Cameron}, {Cosentino},
  {Crossfield}, {Damasso}, {Dressing}, {Dumusque}, {Everett}, {Figueira},
  {Fiorenzano}, {Gonzales}, {Haywood}, {Harutyunyan}, {Hirsch}, {Howell},
  {Johnson}, {Latham}, {Lopez}, {Mayor}, {Micela}, {Molinari}, {Nascimbeni},
  {Pepe}, {Phillips}, {Piotto}, {Rice}, {Sasselov}, {S{\'e}gransan},
  {Sozzetti}, {Udry}, \& {Watson}}]{Malavolta2018}
{Malavolta} L. {et~al.}, 2018, \aj, 155, 107

\bibitem[{{Masuda} {et~al}\mbox{.}(2013){Masuda}, {Hirano}, {Taruya},
  {Nagasawa}, \& {Suto}}]{Masuda2013}
{Masuda} K., {Hirano} T., {Taruya} A., {Nagasawa} M., {Suto} Y., 2013, \apj,
  778, 185

\bibitem[{{Mills} \& {Mazeh}(2017)}]{MillsMazeh2017}
{Mills} S.~M., {Mazeh} T., 2017, \apj, 839, L8

\bibitem[{{Mortier} {et~al}\mbox{.}(2014){Mortier}, {Sousa}, {Adibekyan},
  {Brand{\~a}o}, \& {Santos}}]{Mortier2014}
{Mortier} A., {Sousa} S.~G., {Adibekyan} V.~Z., {Brand{\~a}o} I.~M., {Santos}
  N.~C., 2014, \aap, 572, A95

\bibitem[{{Morton}(2015)}]{Morton2015}
{Morton} T.~D., 2015, {isochrones: Stellar model grid package}. Astrophysics
  Source Code Library

\bibitem[{{Nespral} {et~al}\mbox{.}(2017){Nespral}, {Gandolfi}, {Deeg},
  {Borsato}, {Fridlund}, {Barrag{\'a}n}, {Alonso}, {Grziwa}, {Korth},
  {Albrecht}, {Cabrera}, {Csizmadia}, {Nowak}, {Kuutma}, {Saario},
  {Eigm{\"u}ller}, {Erikson}, {Guenther}, {Hatzes}, {Monta{\~n}{\'e}s
  Rodr{\'\i}guez}, {Palle}, {P{\"a}tzold}, {Prieto-Arranz}, {Rauer}, \&
  {Sebastian}}]{Nespral2017}
{Nespral} D. {et~al.}, 2017, \aap, 601, A128

\bibitem[{{Osborn} {et~al}\mbox{.}(2017){Osborn}, {Santerne}, {Barros},
  {Santos}, {Dumusque}, {Malavolta}, {Armstrong}, {Hojjatpanah}, {Demangeon},
  {Adibekyan}, {Almenara}, {Barrado}, {Bayliss}, {Boisse}, {Bouchy}, {Brown},
  {Cameron}, {Charbonneau}, {Deleuil}, {Delgado Mena}, {D{\'\i}az},
  {H{\'e}brard}, {Kirk}, {King}, {Lam}, {Latham}, {Lillo-Box}, {Louden},
  {Lovis}, {Marmier}, {McCormac}, {Molinari}, {Pepe}, {Pollacco}, {Sousa},
  {Udry}, \& {Walker}}]{Osborn2017}
{Osborn} H.~P. {et~al.}, 2017, \aap, 604, A19

\bibitem[{{Pace}(2013)}]{Pace2013}
{Pace} G., 2013, \aap, 551, L8

\bibitem[{{Paxton} {et~al}\mbox{.}(2011){Paxton}, {Bildsten}, {Dotter},
  {Herwig}, {Lesaffre}, \& {Timmes}}]{Paxton2011}
{Paxton} B., {Bildsten} L., {Dotter} A., {Herwig} F., {Lesaffre} P., {Timmes}
  F., 2011, \apjs, 192, 3

\bibitem[{{Petigura} {et~al}\mbox{.}(2017){Petigura}, {Howard}, {Marcy},
  {Johnson}, {Isaacson}, {Cargile}, {Hebb}, {Fulton}, {Weiss}, {Morton},
  {Winn}, {Rogers}, {Sinukoff}, {Hirsch}, \& {Crossfield}}]{Petigura2017}
{Petigura} E.~A. {et~al.}, 2017, \aj, 154, 107

\bibitem[{{Rauer} {et~al}\mbox{.}(2014){Rauer}, {Catala}, {Aerts},
  {Appourchaux}, {Benz}, {Brandeker}, {Christensen-Dalsgaard}, {Deleuil},
  {Gizon}, {Goupil}, {G{\"u}del}, {Janot-Pacheco}, {Mas-Hesse}, {Pagano},
  {Piotto}, {Pollacco}, {Santos}, {Smith}, {Su{\'a}rez}, {Szab{\'o}}, {Udry},
  {Adibekyan}, {Alibert}, {Almenara}, {Amaro-Seoane}, {Eiff}, {Asplund},
  {Antonello}, {Barnes}, {Baudin}, {Belkacem}, {Bergemann}, {Bihain}, {Birch},
  {Bonfils}, {Boisse}, {Bonomo}, {Borsa}, {Brand{\~a}o}, {Brocato}, {Brun},
  {Burleigh}, {Burston}, {Cabrera}, {Cassisi}, {Chaplin}, {Charpinet},
  {Chiappini}, {Church}, {Csizmadia}, {Cunha}, {Damasso}, {Davies}, {Deeg},
  {D{\'{\i}}az}, {Dreizler}, {Dreyer}, {Eggenberger}, {Ehrenreich},
  {Eigm{\"u}ller}, {Erikson}, {Farmer}, {Feltzing}, {de Oliveira Fialho},
  {Figueira}, {Forveille}, {Fridlund}, {Garc{\'{\i}}a}, {Giommi}, {Giuffrida},
  {Godolt}, {Gomes da Silva}, {Granzer}, {Grenfell}, {Grotsch-Noels},
  {G{\"u}nther}, {Haswell}, {Hatzes}, {H{\'e}brard}, {Hekker}, {Helled},
  {Heng}, {Jenkins}, {Johansen}, {Khodachenko}, {Kislyakova}, {Kley}, {Kolb},
  {Krivova}, {Kupka}, {Lammer}, {Lanza}, {Lebreton}, {Magrin}, {Marcos-Arenal},
  {Marrese}, {Marques}, {Martins}, {Mathis}, {Mathur}, {Messina}, {Miglio},
  {Montalban}, {Montalto}, {Monteiro}, {Moradi}, {Moravveji}, {Mordasini},
  {Morel}, {Mortier}, {Nascimbeni}, {Nelson}, {Nielsen}, {Noack}, {Norton},
  {Ofir}, {Oshagh}, {Ouazzani}, {P{\'a}pics}, {Parro}, {Petit}, {Plez},
  {Poretti}, {Quirrenbach}, {Ragazzoni}, {Raimondo}, {Rainer}, {Reese},
  {Redmer}, {Reffert}, {Rojas-Ayala}, {Roxburgh}, {Salmon}, {Santerne},
  {Schneider}, {Schou}, {Schuh}, {Schunker}, {Silva-Valio}, {Silvotti},
  {Skillen}, {Snellen}, {Sohl}, {Sousa}, {Sozzetti}, {Stello}, {Strassmeier},
  {{\v S}vanda}, {Szab{\'o}}, {Tkachenko}, {Valencia}, {Van Grootel},
  {Vauclair}, {Ventura}, {Wagner}, {Walton}, {Weingrill}, {Werner}, {Wheatley},
  \& {Zwintz}}]{Rauer2014}
{Rauer} H. {et~al.}, 2014, Experimental Astronomy, 38, 249

\bibitem[{{Ricker} {et~al}\mbox{.}(2014){Ricker}, {Winn}, {Vanderspek},
  {Latham}, {Bakos}, {Bean}, {Berta-Thompson}, {Brown}, {Buchhave}, {Butler},
  {Butler}, {Chaplin}, {Charbonneau}, {Christensen-Dalsgaard}, {Clampin},
  {Deming}, {Doty}, {De Lee}, {Dressing}, {Dunham}, {Endl}, {Fressin}, {Ge},
  {Henning}, {Holman}, {Howard}, {Ida}, {Jenkins}, {Jernigan}, {Johnson},
  {Kaltenegger}, {Kawai}, {Kjeldsen}, {Laughlin}, {Levine}, {Lin}, {Lissauer},
  {MacQueen}, {Marcy}, {McCullough}, {Morton}, {Narita}, {Paegert}, {Palle},
  {Pepe}, {Pepper}, {Quirrenbach}, {Rinehart}, {Sasselov}, {Sato}, {Seager},
  {Sozzetti}, {Stassun}, {Sullivan}, {Szentgyorgyi}, {Torres}, {Udry}, \&
  {Villasenor}}]{Ricker2014}
{Ricker} G.~R. {et~al.}, 2014, in \procspie, Vol. 9143, Space Telescopes and
  Instrumentation 2014: Optical, Infrared, and Millimeter Wave, p. 914320

\bibitem[{{Skrutskie} {et~al}\mbox{.}(2006){Skrutskie}, {Cutri}, {Stiening},
  {Weinberg}, {Schneider}, {Carpenter}, {Beichman}, {Capps}, {Chester},
  {Elias}, {Huchra}, {Liebert}, {Lonsdale}, {Monet}, {Price}, {Seitzer},
  {Jarrett}, {Kirkpatrick}, {Gizis}, {Howard}, {Evans}, {Fowler}, {Fullmer},
  {Hurt}, {Light}, {Kopan}, {Marsh}, {McCallon}, {Tam}, {Van Dyk}, \&
  {Wheelock}}]{Skrutskie2006}
{Skrutskie} M.~F. {et~al.}, 2006, \aj, 131, 1163

\bibitem[{{Sneden}(1973)}]{Sneden1973}
{Sneden} C., 1973, \apj, 184, 839

\bibitem[{{Sousa} {et~al}\mbox{.}(2015){Sousa}, {Santos}, {Adibekyan},
  {Delgado-Mena}, \& {Israelian}}]{Sousa2015}
{Sousa} S.~G., {Santos} N.~C., {Adibekyan} V., {Delgado-Mena} E., {Israelian}
  G., 2015, \aap, 577, A67

\bibitem[{{Sozzetti} {et~al}\mbox{.}(2015){Sozzetti}, {Bonomo}, {Biazzo},
  {Mancini}, {Damasso}, {Desidera}, {Gratton}, {Lanza}, {Poretti}, {Rainer},
  {Malavolta}, {Affer}, {Barbieri}, {Bedin}, {Boccato}, {Bonavita}, {Borsa},
  {Ciceri}, {Claudi}, {Gandolfi}, {Giacobbe}, {Henning}, {Knapic}, {Latham},
  {Lodato}, {Maggio}, {Maldonado}, {Marzari}, {Martinez Fiorenzano}, {Micela},
  {Molinari}, {Mordasini}, {Nascimbeni}, {Pagano}, {Pedani}, {Pepe}, {Piotto},
  {Santos}, {Scandariato}, {Shkolnik}, \& {Southworth}}]{Sozzetti2015}
{Sozzetti} A. {et~al.}, 2015, \aap, 575, L15

\bibitem[{{Stassun} \& {Torres}(2018)}]{StassunTorres2018}
{Stassun} K.~G., {Torres} G., 2018, \apj, 862, 61

\bibitem[{{Steffen}(2016)}]{Steffen2016}
{Steffen} J.~H., 2016, \mnras, 457, 4384

\bibitem[{Storn \& Price(1997)}]{Storn1997}
Storn R., Price K., 1997, Journal of Global Optimization, 11, 341

\bibitem[{{Torres} {et~al}\mbox{.}(2011){Torres}, {Fressin}, {Batalha},
  {Borucki}, {Brown}, {Bryson}, {Buchhave}, {Charbonneau}, {Ciardi}, {Dunham},
  {Fabrycky}, {Ford}, {Gautier}, {Gilliland}, {Holman}, {Howell}, {Isaacson},
  {Jenkins}, {Koch}, {Latham}, {Lissauer}, {Marcy}, {Monet}, {Prsa}, {Quinn},
  {Ragozzine}, {Rowe}, {Sasselov}, {Steffen}, \& {Welsh}}]{Torres2011}
{Torres} G. {et~al.}, 2011, \apj, 727, 24

\bibitem[{{Van Eylen} {et~al}\mbox{.}(2016){Van Eylen}, {Albrecht}, {Gandolfi},
  {Dai}, {Winn}, {Hirano}, {Narita}, {Bruntt}, {Prieto-Arranz}, {B{\'e}jar},
  {Nowak}, {Lund}, {Palle}, {Ribas}, {Sanchis-Ojeda}, {Yu}, {Arriagada},
  {Butler}, {Crane}, {Handberg}, {Deeg}, {Jessen-Hansen}, {Johnson}, {Nespral},
  {Rogers}, {Ryu}, {Shectman}, {Shrotriya}, {Slumstrup}, {Takeda}, {Teske},
  {Thompson}, {Vanderburg}, \& {Wittenmyer}}]{VanEylen2016}
{Van Eylen} V. {et~al.}, 2016, \aj, 152, 143

\bibitem[{{von Essen} {et~al}\mbox{.}(2018){von Essen}, {Ofir}, {Dreizler},
  {Agol}, {Freudenthal}, {Hernandez}, {Wedemeyer}, {Parkash}, {Deeg}, {Hoyer},
  {Morris}, {Becker}, {Sun}, {Gu}, {Herrero}, {Tal-Or}, {Poppenhaeger},
  {Mallonn}, {Albrecht}, {Khalafinejad}, {Boumis}, {Delgado-Correal},
  {Fabrycky}, {Janulis}, {Lalitha}, {Liakos}, {Mikolaitis}, {Moyano D'Angelo},
  {Sokov}, {Pakstiene}, {Popov}, {Krushinsky}, {Ribas}, {Rodriguez S.},
  {Rusov}, {Sokova}, {Tautvaisiene}, \& {Wang}}]{vonEssen2018koinet1}
{von Essen} C. {et~al.}, 2018, ArXiv e-prints, arXiv:1801.06191

\bibitem[{{Wang} {et~al}\mbox{.}(2018){Wang}, {Addison}, {Fischer}, {Brewer},
  {Isaacson}, {Howard}, \& {Laughlin}}]{Wang2018}
{Wang} S., {Addison} B., {Fischer} D.~A., {Brewer} J.~M., {Isaacson} H.,
  {Howard} A.~W., {Laughlin} G., 2018, \aj, 155, 70

\bibitem[{{Weiss} {et~al}\mbox{.}(2016){Weiss}, {Deck}, {Sinukoff}, {Petigura},
  {Agol}, {Lee}, {Becker}, {Howard}, {Isaacson}, {Crossfield}, {Fulton}, \&
  {Hirsch}}]{Weiss2016}
{Weiss} L.~M. {et~al.}, 2016, ArXiv e-prints

\bibitem[{{Weiss} \& {Marcy}(2014)}]{Weiss2014}
{Weiss} L.~M., {Marcy} G.~W., 2014, \apjl, 783, L6

\bibitem[{{Weiss} {et~al}\mbox{.}(2013){Weiss}, {Marcy}, {Rowe}, {Howard},
  {Isaacson}, {Fortney}, {Miller}, {Demory}, {Fischer}, {Adams}, {Dupree},
  {Howell}, {Kolbl}, {Johnson}, {Horch}, {Everett}, {Fabrycky}, \&
  {Seager}}]{Weiss2013}
{Weiss} L.~M. {et~al.}, 2013, \apj, 768, 14

\bibitem[{{Winn}(2010)}]{Winn2011_arXiv}
{Winn} J.~N., 2010, ArXiv e-prints

\end{thebibliography}
\bsp

\appendix
\section{Time of transit and durations}\label{sec:tt_t41_appendix}
Central time ($T_{0}$) and duration $T_{14}$ with corresponding error obtained from the fit of each transit, as described in section~\ref{sec:dynamics}.\par

\begin{table}
    \caption{Transit times and durations as described in section~\ref{sec:dynamics}. Full table available in electronic form.}
    \label{tab:tt_t41}
    \begin{tabular}{cccccc}
        \hline
        Transit number  & $T_{0}$ & $\sigma_{T_{0}}$ & $T_{14}$ & $\sigma_{T_{14}}$ & SC/LC  \\
        $^\textrm{(a)}$ & (KBJD$_\textrm{TDB}^\textrm{(b)}$) & (min) & (min) & (min) & \\
        \hline
        \textit{Kepler}-9b & & & & & \\
        0   &    144.24992 &  1.6 &   254.3 &  5.0 & LC \\ 
        1   &    163.48362 &  1.1 &   249.4 &  3.6 & LC \\ 
        3   &    201.95379 &  1.2 &   258.5 &  3.6 & LC \\ 
        ... &    ...       &  ... &     ... &  ... &    \\
        \textit{Kepler}-9c & & & & & \\
        0   &    136.30647 &  1.2 &   288.2 &  5.2 & LC \\ 
        1   &    175.33153 &  1.9 &   289.9 &  4.4 & LC \\ 
        2   &    214.33540 &  1.1 &   279.2 &  4.6 & LC \\ 
        ... &    ...       &  ... &     ... &  ... &    \\
    \hline
    \end{tabular}
    \vspace{0mm}
    \begin{flushleft}
        $^\textrm{(a)}$ Transit number computed with respect to the linear ephemeris in Table~\ref{tab:priors}.\\
        $^\textrm{(b)}$ BJD$_\textrm{TDB}-2454833.0$.
    \end{flushleft}
\end{table}

\section{Radial Velocity measurements with HARPS-N.}\label{sec:rv_appendix}
Radial Velocity with corresponding errors obtained with HARPS-N facility.\par
\begin{table}
    \caption{HARPS-N RVs.}
    \label{tab:rv}
    \begin{tabular}{rrr}
        \hline
        Time  & RV & $\sigma_\textrm{RV}$ \\
        (BJD$_\textrm{TDB}$) & (\ms) & (\ms)  \\
        \hline
        2456783.60953 &   2353.88 &   12.88 \\ 
        2456783.63028 &   2348.14 &   10.00 \\ 
        2456798.63783 &   2338.88 &    8.42 \\ 
        2456798.66056 &   2356.35 &    8.23 \\ 
        2456801.65029 &   2344.83 &    9.25 \\ 
        2456801.66955 &   2333.04 &   10.26 \\ 
        2456813.53960 &   2340.04 &    7.38 \\ 
        2456813.56188 &   2335.65 &    7.33 \\ 
        2456815.54510 &   2331.71 &    9.40 \\ 
        2456815.56727 &   2334.02 &    8.52 \\ 
        2456829.53712 &   2337.37 &   10.28 \\ 
        2456829.55869 &   2326.08 &   10.58 \\ 
        2456831.58087 &   2325.38 &    9.56 \\ 
        2456831.60226 &   2336.21 &   11.00 \\ 
        2456834.50224 &   2311.57 &   15.00 \\ 
        2456845.67515 &   2319.74 &   10.48 \\ 
        2456845.69621 &   2315.50 &   10.85 \\ 
        2456848.44839 &   2342.51 &   15.55 \\ 
        2456848.46990 &   2318.14 &   16.26 \\ 
        2456851.45439 &   2345.14 &    9.33 \\ 
        2456851.47538 &   2326.24 &    9.37 \\ 
        2456853.45667 &   2323.64 &   16.36 \\ 
        2456853.47773 &   2335.83 &   18.50 \\ 
        2456922.38581 &   2314.59 &   13.09 \\ 
        2456922.40526 &   2315.97 &   16.20 \\ 
        2456925.37421 &   2314.17 &   19.65 \\ 
        2456934.36631 &   2338.37 &   11.78 \\ 
        2456934.38747 &   2337.96 &   11.48 \\ 
        2456936.46863 &   2365.09 &   10.49 \\ 
        2456936.49021 &   2348.56 &   10.31 \\
        \hline
    \end{tabular}
\end{table}

\label{lastpage}
\end{document}